\documentclass[gmd,manuscript]{copernicus}
\usepackage{booktabs,threeparttable,seqsplit}

\nolinenumbers

\begin{document}

\title{CLM-JAX v3.2.46: a differentiable, GPU-accelerated land surface model numerically indistinguishable from the Community Land Model 5.0}
\Author[1]{Shijin}{Yuan}
\Author[1]{Zhen}{Wei}
\Author[1]{Tianyu}{Shen}
\Author[1][binmu@tongji.edu.cn]{Bin}{Mu}
\Author[1]{Bo}{Qin}
\Author[2]{Guokun}{Dai}
\affil[1]{School of Computer Science and Technology, Tongji University, Shanghai 201804, China}
\affil[2]{Department of Atmospheric and Oceanic Sciences \& Institute of Atmospheric Sciences, Fudan University, Shanghai 200438, China}
\equalcontrib{1,2}
\correspondence{Bin Mu (\href{mailto:binmu@tongji.edu.cn}{binmu@tongji.edu.cn}).}

\runningtitle{CLM-JAX}
\runningauthor{Shijin Yuan et al.}
\firstpage{1}

\maketitle
\nolinenumbers

\begin{abstract}
The original Fortran implementation of the Community Land Model version 5.0 (CLM5.0) primarily targets CPUs and does not support GPU computation or automatic differentiation.
We present CLM-JAX, a land surface model implemented in Python, based on the CLM5.0 BGC-Crop configuration in the Community Earth System Model version 2.1.3 (CESM2.1.3).
CLM-JAX retains the original subgrid hierarchy, parameterizations and coupling among energy, water, carbon, nitrogen and crop processes, runs on CPUs and GPUs, and supports gradient backpropagation through multi-step time integration.
Using floating-point perturbations of order $10^{-14}$ in the initial state, we construct a stringent ensemble baseline of CLM's own differences and evaluate the numerical consistency of CLM-JAX for 32 variables over five years of continuous global simulation during 2019--2023.
CLM-JAX errors relative to CLM generally lie below this ensemble baseline.
On a single NVIDIA A800 GPU, CLM-JAX requires 0.7692~s per model hour, yielding sustained integration speedups of 4.04 and 2.57 over CLM compiled with gfortran and ifort, respectively, on 64 CPU cores.
In the gradient test, the objective function is the first-layer soil temperature at step 100 (50~h), and its gradient with respect to the initial soil temperature field of all model layers is obtained by backpropagation through the 100 time steps.
With converged central finite differences as references, 95.79\% of the 84,254 nonzero references have relative errors no greater than 1\% and 98.22\% have matching signs, with median and 90th-percentile relative errors of $5.862\times10^{-7}$\% and $3.578\times10^{-4}$\%, respectively.
The differentiability of CLM-JAX provides a basis for gradient-based applications, such as end-to-end parameter optimization and four-dimensional variational data assimilation.

\end{abstract}

\introduction
\label{sec:introduction}

Land surface models (LSMs) simulate the exchanges of energy, water and carbon between the land surface and the atmosphere, together with terrestrial nitrogen cycling, and are core components of Earth system models (ESMs). As LSMs have incorporated vegetation physiology, nutrient limitation and land-use change, both their computational cost and the degree of coupling among processes have increased \citep{pitman2003evolution,fisher2020perspectives}. The Community Land Model version 5.0 (CLM5.0), the land component of the Community Earth System Model version 2 (CESM2), is one of the most comprehensive and widely used LSMs, representing biogeophysical, biogeochemical and crop processes within a nested subgrid hierarchy \citep{lawrence2019clm5}. This breadth of processes brings many uncertain parameters, and constraining them with observations or quantifying their effects through large ensembles requires many costly simulations \citep{dagon2020machine,kennedy2025parameter}.

Many LSM applications would benefit from faster simulations and from gradient information. CLM is written in Fortran and runs primarily on CPUs using MPI. OpenACC has been used to port parts of CLM and the E3SM Land Model (ELM) to GPUs \citep{wang2014openacc,schwartz2022openacc}, but this approach retains the Fortran code and does not by itself provide automatic differentiation. Differentiable modeling expresses physical processes and their numerical solutions with operations that support automatic differentiation, so that parameter estimation, sensitivity analysis and the joint optimization of physical models and neural networks can use gradient information \citep{gelbrecht2023differentiable,shen2023differentiable}. JAX provides automatic differentiation, just-in-time compilation and parallel computation on CPUs and GPUs in Python \citep{bradbury2018jax}.

Python implementations, many of them differentiable, have been developed for land surface and ecosystem models, mostly for simplified models or individual modules. Examples include the simplified ELM (sELM) used to build machine-learning surrogates \citep{lu2019efficient}, the leaf photosynthesis module of the Functionally Assembled Terrestrial Ecosystem Simulator (FATES) in diffEcosys \citep{aboelyazeed2023differentiable}, and the carbon and water dynamics of the Data Assimilation Linked Ecosystem Carbon (DALEC) model in JAX, later extended to global parameter estimation in DifferLand \citep{fang2024exploring,fang2026differentiable}. Models with multiple coupled processes include the multilayer canopy model CanVeg in JAX \citep{jiang2025jax}, NoahPy for soil thermohydrology in PyTorch \citep{tian2026noahpy}, ADELM v1.0 with canopy, soil and snow processes in PyTorch \citep{jiang2026adelm}, CoLM Python v1.0 for the Common Land Model (CoLM) \citep{colm_python_v1}, and legoESM, a differentiable Earth system model built in JAX \citep{gentine2026legoesm}. For CLM itself, large language models (LLMs) have been used to translate the leaf photosynthesis module \citep{zhou2024proof} and the CLM-ml-v2 multilayer canopy model \citep{lahlou2026systematic} into JAX.

These studies demonstrate the potential of differentiable and Python-based land surface modeling, but they mainly target simplified models, individual processes or models with fewer coupled processes than CLM5.0, and those related to CLM address individual modules or its multilayer canopy branch. To our knowledge, no previous study reimplements the full set of coupled biogeophysical and biogeochemical processes in the CLM5.0 BGC-Crop configuration in a differentiable, GPU-capable framework, or evaluates the numerical consistency of such a reimplementation with CLM in multiyear global simulations. Module-level tests cannot replace this evaluation, because the subgrid hierarchy, processes on multiple timescales and carbon--nitrogen cycling interact in the full model, and some annually updated accumulators affect subsequent states only in simulations spanning year boundaries.

To address this gap, we develop CLM-JAX, a land surface model implemented in Python using JAX, based on the CLM5.0 BGC-Crop configuration in CESM2.1.3. Because floating-point rounding differences can be amplified during continuous integration, its consistency with CLM is assessed against CLM's own response to small perturbations. We adapt the ensemble consistency testing method of \citet{baker2015pycect} for use in the land model, construct an ensemble by applying floating-point perturbations of order $10^{-14}$~K to the initial soil temperature, define a baseline of CLM's own numerical differences, and compare the differences between CLM-JAX and CLM with the baseline of CLM's own numerical differences.

The main contributions of this study are as follows:
\begin{enumerate}
  \item CLM-JAX retains the full subgrid hierarchy, parameterizations and process coupling of the original configuration, reads CESM input files for global simulations and runs on CPUs and GPUs.
  \item Over five years of continuous global simulation, CLM-JAX errors relative to CLM for 32 variables generally lie below the ensemble baseline of CLM's own differences.
  \item On a single NVIDIA A800 GPU, CLM-JAX integrates 4.04 and 2.57 times faster than CLM compiled with gfortran and ifort, respectively, on 64 CPU cores, with both models in double precision.
  \item CLM-JAX supports gradient backpropagation through multi-step time integration; after 100 time steps, its directional derivatives agree with converged finite differences to within 1\% in 95.79\% of comparisons.
\end{enumerate}

The remainder of this paper is organized as follows. Section~2 describes CLM5.0, the implementation of CLM-JAX and the experimental design, Section~3 presents the results, and Sections~4 and~5 provide the discussion and conclusions, respectively.

\section{Materials and methods}

\subsection{Community Land Model version 5.0}
\label{sec:clm5-overview}

CLM5.0 can be run coupled to an atmospheric model within CESM2 or with prescribed atmospheric forcing \citep{lawrence2019clm5}. This study uses the I2000Clm50BgcCrop component set, which runs CLM5.0 in the BGC-Crop configuration. In this configuration, vegetation carbon and nitrogen, phenology and crop growth are simulated prognostically. Each gridcell is divided into a nested hierarchy of landunits, columns and patches (Fig.~S1 in the Supplement; \citealp{clm2018technical}). Landunits represent natural vegetation, crops, lakes, urban areas and glaciers. Columns carry the soil and snow states, with up to 12 snow layers and 25 ground layers, the upper 20 of which are soil. Patches represent plant functional types (PFTs). On the natural vegetated landunit, all PFTs share a single soil column and compete for soil water and nutrients, whereas each crop type has separate rainfed and irrigated columns.

Biogeophysical processes include canopy interception, radiative transfer, surface energy balance, evapotranspiration, soil and snow heat transfer, and soil water movement. Key schemes include the Medlyn stomatal conductance model \citep{medlyn2011stomatal}, the Leaf Utilization of Nitrogen for Assimilation (LUNA) scheme for photosynthetic capacity \citep{ali2016luna}, plant hydraulic stress \citep{kennedy2019hydraulics}, an adaptive Richards equation solver for soil water, and the Snow, Ice, and Aerosol Radiative (SNICAR) model for snow radiative transfer \citep{flanner2007snow}. Biogeochemical processes govern carbon and nitrogen pools in vegetation, litter and soil, including CENTURY-based vertically resolved decomposition \citep{parton1988dynamics,koven2013vertical}, plant nitrogen uptake with the Fixation and Uptake of Nitrogen (FUN) scheme \citep{fisher2010nitrogen,shi2016nitrogen} and flexible carbon-to-nitrogen ratios, and methane fluxes. The crop module simulates planting, phenology, irrigation, fertilization and harvest. These processes span timescales from the half-hourly time step to annual updates of accumulated quantities \citep{lawrence2019clm5}. Table~S1 in the Supplement summarizes the major processes and their order within a time step.

\subsection{Implementation of CLM-JAX}
\label{sec:clm-jax-implementation}

\begin{figure}[!htbp]
  \centering
  \includegraphics[width=1.0\linewidth,height=0.79\textheight,keepaspectratio]{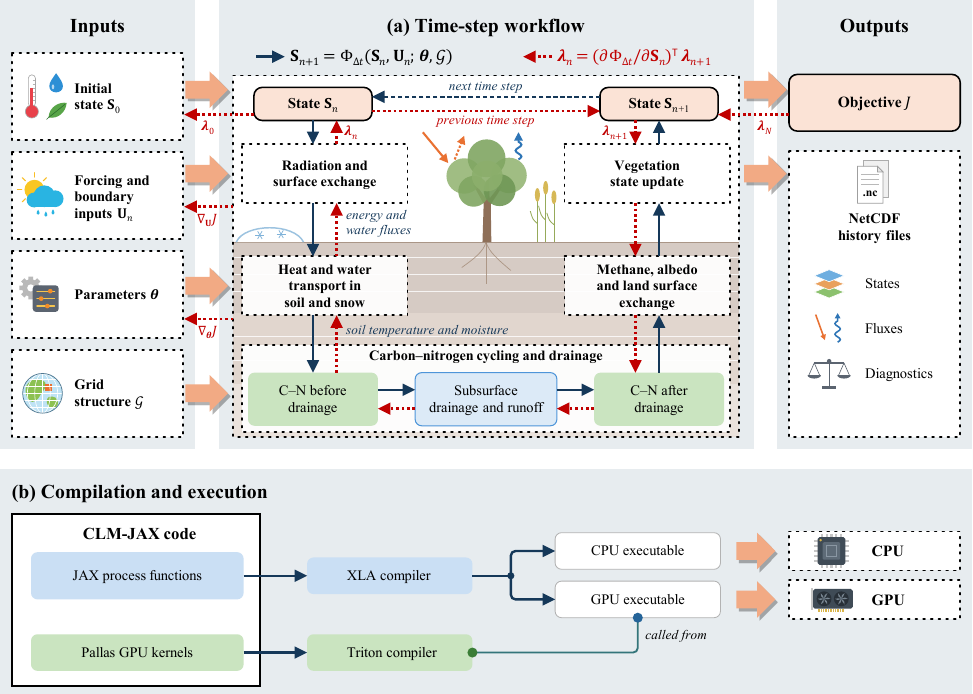}
  \caption{CLM-JAX workflow and execution. (a) Forward and backward passes through a model time step. Starting from the initial state $\mathbf{S}_0$, the forward pass (dark blue arrows) advances the state $\mathbf{S}_n$ to $\mathbf{S}_{n+1}$ using forcing and boundary inputs $\mathbf{U}_n$, parameters $\boldsymbol{\theta}$ and grid structure $\mathcal{G}$, as in Eq.~\eqref{eq:clm-jax-state-transition}. Carbon and nitrogen calculations are performed before and after subsurface drainage. History files record states, fluxes and diagnostics after the vegetation state update. In the backward pass (red dotted arrows), the adjoint $\boldsymbol{\lambda}_N=\nabla_{\mathbf{S}_N}J$ obtained from the objective $J$ is propagated backward in time and, within each time step, through the processes in reverse order, as in Eq.~\eqref{eq:clm-jax-adjoint}. The resulting $\boldsymbol{\lambda}_0$ is the gradient of $J$ with respect to the initial state, and gradients with respect to the continuous forcing and the parameters can be computed in the same way. (b) Compilation and execution. XLA compiles JAX process functions into CPU or GPU executables. Selected GPU kernels are written in Pallas, compiled by Triton, and called from the GPU executable.}
  \label{fig:clm-jax}
\end{figure}

Fig.~\ref{fig:clm-jax} gives an overview of CLM-JAX. The forward computation within a model time step and the backward propagation of gradients through it (Fig.~\ref{fig:clm-jax}a) are described in Sections~\ref{sec:clm-jax-state-transition} and~\ref{sec:clm-jax-differentiation}, and the compilation and execution of the model on CPUs and GPUs (Fig.~\ref{fig:clm-jax}b) are described in Section~\ref{sec:clm-jax-execution}.

\subsubsection{Model framework and subgrid representation}
\label{sec:clm-jax-framework}

CLM-JAX implements the processes enabled in the I2000Clm50BgcCrop component set and retains their original formulations. It preserves the CLM hierarchy of gridcells, landunits, columns and patches. Subgrid elements are stored as one-dimensional arrays in the same order as in CLM, and index arrays link each patch to its column, landunit and gridcell. Soil and snow variables are stored by column and layer, and vegetation variables by patch. Whereas the Fortran model distributes gridcells across MPI tasks, CLM-JAX holds all land gridcells on a single device and processes them with vectorized array operations. All calculations use double precision.

CLM uses filters, which list the indices of active elements, to select the columns and patches on which each process operates, such as soil, snow-covered, lake or crop elements. CLM-JAX instead uses boolean masks and fixed-size index arrays, because just-in-time compilation in JAX requires array shapes to be known in advance. Some processes gather the selected elements into compact arrays, compute on them and scatter the results back, avoiding computation on inactive elements. Subgrid fluxes and states are aggregated to landunits and gridcells using the CLM area weights, following the same summation order as the original model to limit rounding differences arising from the non-associativity of floating-point addition.

\subsubsection{Time stepping and process coupling}
\label{sec:clm-jax-state-transition}

The model state is stored in an immutable JAX PyTree, a nested collection of arrays that includes water storage, temperature, carbon and nitrogen pools, and persistent auxiliary quantities such as phenological accumulators and variables needed for lagged feedbacks. A complete time step is written as
\begin{equation}
  \mathbf{S}_{n+1}
  = \Phi_{\Delta t}\!\left(\mathbf{S}_n,\mathbf{U}_n;
    \boldsymbol{\theta},\mathcal{G}\right),
  \label{eq:clm-jax-state-transition}
\end{equation}
where $\mathbf{S}_n$ is the persistent state at step $n$, $\mathbf{U}_n$ contains atmospheric forcing, external coupling inputs and time information, $\boldsymbol{\theta}$ denotes model parameters, and $\mathcal{G}$ describes the grid and subgrid topology. The operator $\Phi_{\Delta t}$ advances the state over $\Delta t=1800$~s using a no-leap calendar. Because the state is immutable, each process returns updated values instead of modifying them in place, so dependencies that the Fortran code passes through shared module variables become explicit inputs and outputs.

The process sequence follows the CLM driver (Fig.~\ref{fig:clm-jax}a). Radiation and surface exchange calculations provide the energy and water fluxes that drive heat and water transport in soil and snow. The updated soil temperature and moisture then regulate the carbon and nitrogen calculations, which are split around subsurface drainage as in CLM so that nitrogen losses through leaching and runoff use the hydrological results of the same time step. Methane fluxes, surface albedo for the next time step and land surface exchange are calculated next, and water, energy, carbon and nitrogen balances are checked at the corresponding process boundaries. The vegetation state is updated at the end of the time step, after which states, fluxes and diagnostics are written to the history files and $\mathbf{S}_{n+1}$ is passed to the next time step.

\subsubsection{Differentiable numerical implementation}
\label{sec:clm-jax-differentiation}

Equation~(\ref{eq:clm-jax-state-transition}) expresses one time step as the state transition operator $\Phi_{\Delta t}$, and a simulation applies this operator repeatedly from the initial state $\mathbf{S}_0$. Following the process sequence within a time step, $\Phi_{\Delta t}$ is itself a composition of process operators,
\begin{equation}
  \Phi_{\Delta t} = \phi_K \circ \phi_{K-1} \circ \cdots \circ \phi_1,
  \label{eq:clm-jax-composition}
\end{equation}
where $\phi_k$ denotes the $k$th process in the time step and its output includes the updated state and the fluxes passed to subsequent processes. All processes are implemented with JAX array operations and control flow, which retain the dependence of the model state on initial conditions, continuous forcing and parameters across time steps. The gradient of an objective function of the simulated states with respect to these quantities can therefore be computed by backpropagation through time (Fig.~\ref{fig:clm-jax}a).

The objective $J$ is a scalar computed from the state after $N$ steps; in this study, it is the area-weighted gridcell value of the first-layer soil temperature. The adjoint $\boldsymbol{\lambda}_n = \nabla_{\mathbf{S}_n} J$, the gradient of $J$ with respect to $\mathbf{S}_n$, has the same dimension as $\mathbf{S}_n$. Backpropagation first obtains $\boldsymbol{\lambda}_N$ from the objective and then propagates the adjoint backward as
\begin{equation}
  \boldsymbol{\lambda}_n
  = \left(\frac{\partial \Phi_{\Delta t}}{\partial \mathbf{S}_n}\right)^{\!\mathsf{T}} \boldsymbol{\lambda}_{n+1}
  = \left(\frac{\partial \phi_1}{\partial \mathbf{z}^{(0)}}\right)^{\!\mathsf{T}}
    \left(\frac{\partial \phi_2}{\partial \mathbf{z}^{(1)}}\right)^{\!\mathsf{T}}
    \cdots
    \left(\frac{\partial \phi_K}{\partial \mathbf{z}^{(K-1)}}\right)^{\!\mathsf{T}}
    \boldsymbol{\lambda}_{n+1},
  \qquad n = N-1,\ldots,0,
  \label{eq:clm-jax-adjoint}
\end{equation}
where $\mathbf{z}^{(k-1)}$ is the input to process $k$ and $\mathbf{z}^{(0)} = \mathbf{S}_n$. The adjoint thus travels backward in time and, within each time step, passes through the processes in the reverse of their forward order (Fig.~\ref{fig:clm-jax}a).

In this study, the control vector $\mathbf{x}$ consists of the initial temperatures of all ground layers in the active soil and crop columns. Because these temperatures are components of $\mathbf{S}_0$, the gradient $\nabla_{\mathbf{x}} J$ consists of the corresponding components of $\boldsymbol{\lambda}_0$. A single backward pass yields all of these components, and its cost does not depend on the dimension of $\mathbf{x}$.

\subsubsection{Computational execution and data interfaces}
\label{sec:clm-jax-execution}

The CPU and GPU versions share the same process interfaces and state definitions. As shown in Fig.~\ref{fig:clm-jax}b, XLA compiles the JAX process functions into CPU or GPU executables, and selected GPU kernels are written in Pallas and compiled by Triton \citep{jaxPallas}. Each process runs in parallel across columns and patches, while vertical calculations within each column remain sequential. Related calculations are combined within a kernel to reduce intermediate memory transfers, and compiled functions and static data are kept on the device and reused across time steps.

Input routines read the CESM case settings, surface dataset, parameter file and restart files, so CLM-JAX uses the input files of a CESM case directly. Atmospheric forcing is prepared with the same temporal and spatial mappings and downscaling as in CESM. MOSART river exchanges are supplied as external boundary inputs at coupling intervals. History output is written to NetCDF files that retain CLM variable names, units and subgrid organization, so it can be compared directly with CLM output, and the final model state can be used to restart the simulation.

\subsection{Experimental setup and data}
\label{sec:experimental-setup-data}

CLM and CLM-JAX use the same configuration, the I2000Clm50BgcCrop component set of CESM2.1.3 on the f09\_g17 grid (approximately $0.9^{\circ}\times1.25^{\circ}$). Following the I2000 definition, land cover, atmospheric CO$_2$ and nitrogen deposition are fixed at year-2000 levels. Irrigation is applied to irrigated crop columns without limitation by river water availability, and atmospheric methane concentration is prescribed.

Atmospheric forcing is taken from version 3.0 of WFDE5, which applies the WATCH Forcing Data methodology to ERA5 through spatial aggregation and bias adjustment \citep{cucchi2020wfde5}. Version 3.0 provides hourly fields over land at $0.5^{\circ}$ resolution for 1979--2024, including near-surface wind speed, air temperature, specific humidity, surface pressure, downward shortwave and longwave radiation, rainfall and snowfall.

A spin-up simulation with CLM from 1979 to 2018 starts from the default CESM2.1.3 initial conditions for this configuration and allows the land state to adjust to the WFDE5 forcing. CLM-JAX and CLM then start from the same restart state on 1 January 2019 and run continuously to the end of 2023, with CLM providing the reference simulation. Because both models share this initial state, slowly evolving carbon and nitrogen pools do not need to be in full equilibrium for the consistency evaluation.

\subsection{Validations}

\subsubsection{Validation of numerical consistency}
\label{sec:numerical-consistency-validation}

CLM-JAX is evaluated against the unperturbed reference CLM simulation, which shares its initial state, atmospheric forcing, boundary inputs, parameters and time step. We evaluate 32 variables covering land states, surface fluxes, vegetation and carbon cycling using six-hourly instantaneous outputs. Table~S2 in the Supplement lists their definitions and units.

At each output time, values are paired by land gridcell and, for soil variables, by layer. Only pairs valid in both simulations are retained, and all pairs are weighted equally. For $N$ valid pairs, let $x_k$ and $r_k$ denote the CLM-JAX and reference CLM values. We calculate the root mean square error (RMSE), normalized root mean square error (nRMSE), Nash--Sutcliffe efficiency (NSE) and Pearson correlation coefficient (CC) as
\begin{equation}
  \mathrm{RMSE}
  = \left[\frac{1}{N}\sum_{k=1}^{N}(x_k-r_k)^2\right]^{1/2},
  \label{eq:consistency-rmse}
\end{equation}
\begin{equation}
  \mathrm{nRMSE}
  = \frac{\mathrm{RMSE}}
  {\left[\frac{1}{N}\sum_{k=1}^{N}r_k^2\right]^{1/2}}
  \times 100\%,
  \label{eq:consistency-nrmse}
\end{equation}
\begin{equation}
  \mathrm{NSE}
  = 1-\frac{\sum_{k=1}^{N}(x_k-r_k)^2}
  {\sum_{k=1}^{N}(r_k-\bar{r})^2},
  \label{eq:consistency-nse}
\end{equation}
\begin{equation}
  \mathrm{CC}
  = \frac{\sum_{k=1}^{N}(x_k-\bar{x})(r_k-\bar{r})}
  {\left[\sum_{k=1}^{N}(x_k-\bar{x})^2
  \sum_{k=1}^{N}(r_k-\bar{r})^2\right]^{1/2}},
  \label{eq:consistency-cc}
\end{equation}
where $\bar{x}$ and $\bar{r}$ are the means over the same $N$ pairs. RMSE retains the units of the variable, and nRMSE expresses it as a percentage of the reference root mean square. NSE measures squared error relative to the variance of the reference, while CC measures linear association. A metric is undefined when its denominator is zero.

The construction of the baseline of CLM's own numerical differences draws on the ensemble consistency testing method of \citet{baker2015pycect}. Ten additional CLM runs are performed, each with a different additive perturbation of order $10^{-14}$~K to the initial soil temperature and otherwise identical to the reference run. For each variable $v$ and output time $t$, the four metrics are calculated for each of the 45 member pairs $(i,j)$ with $i<j$, using member $i$ as the reference. For metric $M\in\{\mathrm{RMSE},\mathrm{nRMSE},\mathrm{NSE},\mathrm{CC}\}$, the baseline is the median of the 45 pairwise values,
\begin{equation}
  B_{v,t}^{M}
  = \underset{1\leq i<j\leq 10}{\operatorname{median}}
    \left\{M_{v,t}^{(i,j)}\right\}.
  \label{eq:consistency-baseline}
\end{equation}
Because the members differ from the reference only by round-off-level perturbations of $10^{-14}$, the ensemble members defining this baseline should, in principle, be regarded as identical, indistinguishable predictions. Nevertheless, numerical differences arising from floating-point rounding between these predictions can still be amplified by the chaotic system during computation. During development, we also found that any error in the CLM-JAX implementation led to simulation errors substantially above this baseline. Comparing the different error metrics against this baseline throughout five years of continuous integration provides a stringent test of numerical consistency. Integration across year boundaries also tests the numerical consistency of annual updates and slowly varying processes.

Each CLM-JAX metric is compared with its baseline and with the spread among the member pairs.

Three additional comparisons examine time-averaged and sampled agreement. First, each simulation is averaged over 2019--2023, and RMSE and nRMSE are calculated across the resulting mean fields. Second, the five-year means are grouped by PFT group or CLM column type and by $10^{\circ}$ latitude band using area weights, with RMSE and nRMSE calculated from patch- and column-level means. Third, to compare individual six-hourly values over their full range, 100,000 paired values are sampled at random across time and land gridcells for each variable, and all four metrics are calculated from these samples. For soil variables, these comparisons use the first soil layer.

\subsubsection{Evaluation of computational performance}
\label{sec:performance-evaluation}

Performance is measured over the continuous five-year integrations from 2019 to 2023, comprising 87,600 time steps, with identical configuration, initial state and forcing for both models. Both models use double precision. All experiments are performed on a single server with 64 AMD EPYC 7542 CPU cores and one NVIDIA A800 GPU. CLM is run on the server's CPU cores, using 1, 2, 4, 8, 16, 32 and 64 cores and one MPI process per core, compiled with GNU Fortran 8.3.0 and Intel ifort 2021.10.0 at optimization level -O2. CLM-JAX is run on the GPU of the same server using JAX~0.10.2, jaxlib~0.10.2 and the CUDA~12.9 runtime.

Elapsed integration time excludes initialization, file output and JAX compilation. For CLM, it is the land component run time reported in the CESM timing output; for CLM-JAX, it is the sum of the per-step wall-clock times recorded by the model driver. The computational cost is expressed as the mean wall-clock time per model hour,
\begin{equation}
  \tau = \frac{T_{\mathrm{int}}}{H},
  \label{eq:integration-cost}
\end{equation}
where $T_{\mathrm{int}}$ is the elapsed integration time in seconds and $H$ is the simulated duration in hours, equal to 43,800 for these tests. The GPU speedup relative to CLM on $p$ CPU cores and the CPU parallel speedup are
\begin{equation}
  S_{\mathrm{GPU}}(p)
  = \frac{\tau_{\mathrm{CLM}}(p)}{\tau_{\mathrm{JAX}}},
  \qquad
  S_{\mathrm{CPU}}(p)
  = \frac{\tau_{\mathrm{CLM}}(1)}{\tau_{\mathrm{CLM}}(p)},
  \label{eq:performance-speedup}
\end{equation}
where $\tau_{\mathrm{CLM}}(p)$ and $\tau_{\mathrm{JAX}}$ are the costs per model hour for CLM on $p$ CPU cores and CLM-JAX on one GPU, respectively. Speedups are calculated separately for each compiler.

Mean CPU utilization is calculated by summing utilization across all model MPI processes, taking a time-weighted mean during integration and dividing by the CPU core count. GPU utilization is sampled every 5~s using nvidia-smi, and its mean is calculated from the samples recorded during integration.

\subsubsection{Validation of CLM-JAX differentiability}
\label{sec:gradient-validation}

Gradient correctness is tested with the same model configuration, initial state and forcing, integrating continuously for 100 time steps, or 50~h of model time, from 1 January 2019. Each gridcell is evaluated as a separate sample. The control vector $\mathbf{x}_g$ of gridcell $g$ consists of the initial temperatures of the 25 ground layers in every active soil and crop column of the gridcell, so a gridcell containing $n_g$ such columns has a control vector with $25n_g$ components. All other initial states, model parameters and forcing are held fixed. The objective function of each gridcell is the first-layer soil temperature at step 100, aggregated to the gridcell following the definition of the native CLM \texttt{TSOI} output:
\begin{equation}
  J_g(\mathbf{x}_g)
  = \frac{\displaystyle\sum_{c\in\mathcal{C}_g} a_c\,
      T_{c,1}^{(100)}(\mathbf{x}_g)}
    {\displaystyle\sum_{c\in\mathcal{C}_g}a_c},
  \label{eq:gradient-gridcell-objective}
\end{equation}
where $\mathcal{C}_g$ is the set of active soil and crop columns in the gridcell, $a_c$ is the area fraction of column $c$ within its gridcell, and $T_{c,1}^{(100)}$ is the temperature of the first ground layer at the end of the full integration. Gridcells with zero soil and crop area have missing \texttt{TSOI} output and are excluded from both the objective and the gradient error comparison.

Under the prescribed external boundary conditions used in this experiment, gridcells evolve independently of one another. Taking the sum of all valid gridcell objectives, $L=\sum_{g\in\mathcal{V}}J_g$, as the scalar target for backpropagation therefore yields the gradient of each gridcell, $\mathbf{q}_g=\nabla_{\mathbf{x}_g}L=\nabla_{\mathbf{x}_g}J_g$, where $\mathcal{V}$ is the set of gridcells with valid \texttt{TSOI}. Backpropagation spans the full 100-step coupled integration.

Following the directional-derivative comparison of JAX-CanVeg and the full-field random-direction test of WRF-JAX \citep{jiang2025jax,yuan2026wrfjax}, eight random directions $\mathbf{v}^{(d)}$ with independent standard Gaussian components are generated, where $d=1,\dots,8$ indexes the direction. Each direction is normalized to unit root-mean-square over the complete control field, which comprises all 25 ground layers of the 77,229 active soil and crop columns. The components $\mathbf{v}^{(d)}_g$ belonging to gridcell $g$ are extracted without renormalization. Each perturbation is applied to the entire initial temperature field at once. Separate CLM-JAX forward integrations are performed for the positive and negative perturbations, and the gridcell objectives at step 100 are used to compute central finite differences. These are compared with the projection of the backpropagated gradient onto the same direction:
\begin{equation}
  \begin{aligned}
    D^{\mathrm{FD}}_{g,d}(h)
      &= \frac{J_g(\mathbf{x}_g+h\mathbf{v}^{(d)}_g)
                 -J_g(\mathbf{x}_g-h\mathbf{v}^{(d)}_g)}{2h},\\
    D^{\mathrm{AD}}_{g,d}
      &= \mathbf{q}_g^{\mathsf{T}}\mathbf{v}^{(d)}_g.
  \end{aligned}
  \label{eq:gradient-directional-comparison}
\end{equation}
The one-sided perturbation amplitude $h$ takes 11 values: $0.1$, $0.03$, $0.01$, $0.003$, $0.001$, $10^{-4}$, $10^{-5}$, $10^{-6}$, $10^{-7}$, $10^{-8}$ and $10^{-9}$~K. The positive and negative perturbations along the eight directions at all amplitudes require 176 forward integrations.

Finite-difference references are selected separately for each gridcell and direction, using forward results only. All nine windows of three adjacent amplitudes are examined. Let $\mathcal{D}$ denote the set of three finite-difference estimates in a window and $m$ their median. A window is considered converged when all three estimates are finite and
\begin{equation}
  \max\mathcal{D}-\min\mathcal{D}\leq 0.01\lvert m\rvert .
  \label{eq:gradient-fd-convergence}
\end{equation}
Among converged windows with a nonzero median, the window with the smallest relative spread $(\max\mathcal{D}-\min\mathcal{D})/\lvert m\rvert$ is selected, and its median is taken as the reference value $R_{g,d}$; ties are resolved in favor of larger amplitudes. When the median is zero, the criterion is satisfied only if all three estimates are zero. Such all-zero windows are used only when no converged window with a nonzero median exists, in which case the all-zero window with the largest amplitudes is selected and the reference is set to zero. If none of the nine windows satisfies the criterion, the finite difference for that gridcell and direction is classified as non-converged.

For nonzero references, the percentage relative error is defined as
\begin{equation}
  E_{g,d}=100\,
    \frac{\lvert D^{\mathrm{AD}}_{g,d}-R_{g,d}\rvert}
         {\lvert R_{g,d}\rvert}.
  \label{eq:gradient-relative-error}
\end{equation}
A relative error of at most 1\% is used as the agreement criterion, and the 10th, 25th, 50th, 75th and 90th percentiles of the relative error are reported. The percentiles are computed over all nonzero converged references, including comparisons with relative errors above 1\%. Zero references are evaluated separately with an absolute error criterion, using the tolerance
\begin{equation}
  \delta_{g,d}=\min_{h\in\mathcal{H}^{*}_{g,d}}
    \frac{\max\{\operatorname{ulp}(J_g^{+}(h)),
                 \operatorname{ulp}(J_g^{-}(h))\}}{2h},
  \label{eq:gradient-zero-tolerance}
\end{equation}
where $\mathcal{H}^{*}_{g,d}$ is the set of three selected perturbation amplitudes, $J_g^{\pm}(h)$ are the objective values for the corresponding positive and negative perturbations, and $\operatorname{ulp}$ denotes the spacing of double-precision floating-point numbers at the objective value (unit in the last place). For a zero reference, the comparison satisfies the absolute error tolerance when $\lvert D^{\mathrm{AD}}_{g,d}\rvert\leq\delta_{g,d}$.

To characterize how the finite differences converge with perturbation amplitude, the relative deviation $\epsilon_{g,d}(h)$ of the central difference at each amplitude is computed relative to the directional projection of the backpropagated gradient, together with the round-off error bound $\beta_{g,d}(h)$ associated with a single ulp:
\begin{equation}
  \begin{aligned}
    \epsilon_{g,d}(h)
      &= 100\,\frac{\lvert D^{\mathrm{FD}}_{g,d}(h)-D^{\mathrm{AD}}_{g,d}\rvert}
                   {\lvert D^{\mathrm{AD}}_{g,d}\rvert},\\
    \beta_{g,d}(h)
      &= 100\,\frac{\max\{\operatorname{ulp}(J_g^{+}(h)),
                          \operatorname{ulp}(J_g^{-}(h))\}}
                   {2h\,\lvert D^{\mathrm{AD}}_{g,d}\rvert}.
  \end{aligned}
  \label{eq:gradient-amplitude-sweep}
\end{equation}
The relative spread of the three finite-difference estimates in the selected window is denoted $S_{g,d}=100\,(\max\mathcal{D}-\min\mathcal{D})/\lvert R_{g,d}\rvert$; by Eq.~\eqref{eq:gradient-fd-convergence}, $S_{g,d}\leq1\%$.

Valid gridcells are divided into two classes according to their thermal state at the initial time. Gridcells in which no active soil or crop column has snow layers and all 25 ground layers of these columns are above freezing (273.15~K) are classified as snow-free, unfrozen; all other gridcells are classified as snow-covered or frozen.

\section{Results}
\subsection{Numerical consistency with CLM}

We assess whether CLM-JAX preserves the numerical behavior of CLM during continuous integration. Of the 32 variables evaluated, eight are shown in the main text: soil temperature (TSOI), volumetric soil water content (H2OSOI), total evapotranspiration (ET), snow water equivalent (H2OSNO), sensible heat flux (FSH), total projected leaf area index (TLAI), gross primary production (GPP) and net primary production (NPP); the other 24 are presented in the Supplement.

Figs.~\ref{fig:key8_rmse} and~\ref{fig:key8_nrmse} clearly show that, for all eight variables, the RMSE and nRMSE between CLM-JAX and CLM generally lie below the baseline of CLM's own numerical differences over 2019--2023. Figs.~S2 and~S3 in the Supplement show the NSE and CC time series for the same eight variables. The same pattern is evident for the remaining 24 variables in Figs.~S4--S9 in the Supplement. Figs.~S10--S15 in the Supplement provide the corresponding NSE and CC time series for these 24 variables.

\begin{figure}[!htbp]
  \centering
  \centerline{\includegraphics[width=1.0\textwidth,height=0.79\textheight,keepaspectratio]{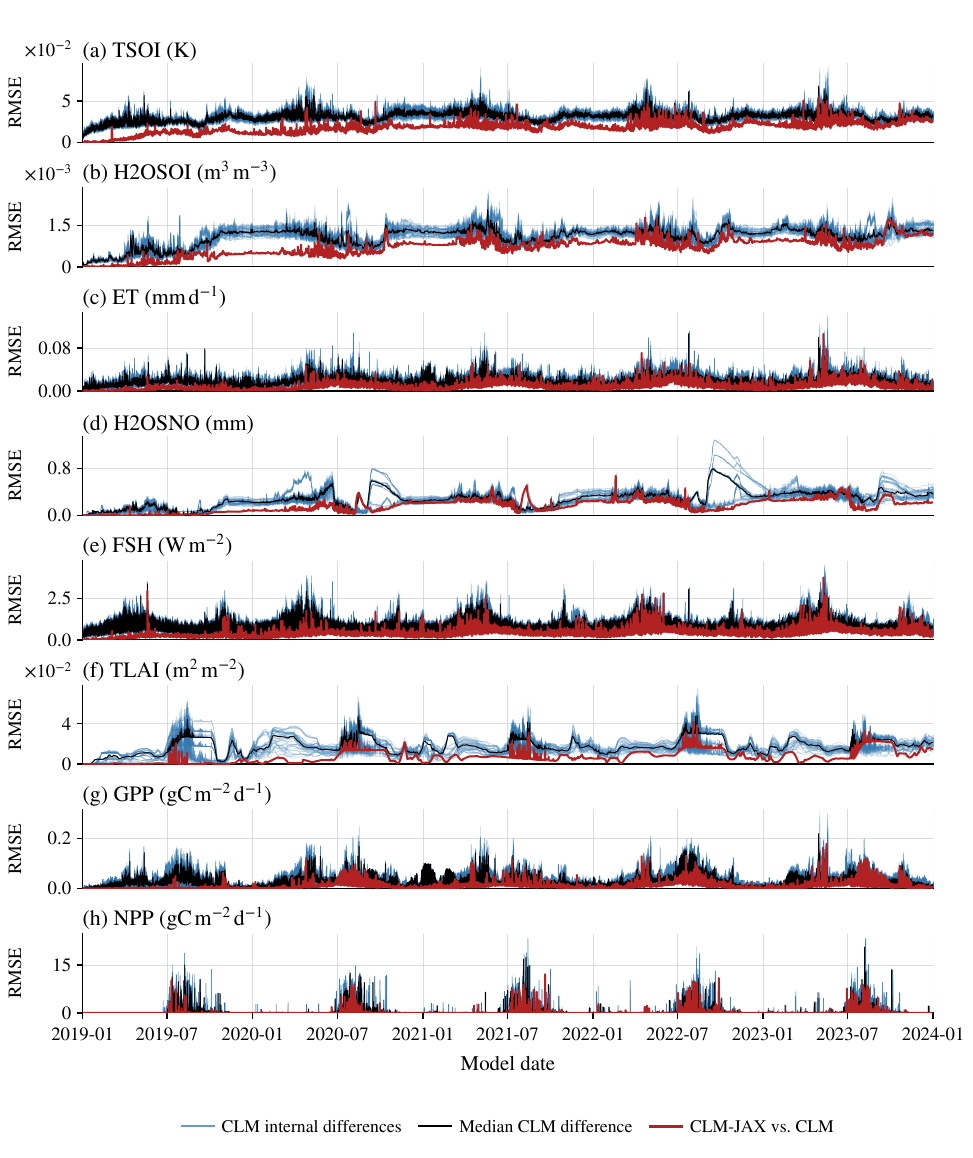}}
  \caption{Six-hourly RMSE of eight variables during 2019--2023. Blue lines show the 45 pairwise
  differences among ten CLM runs with perturbed initial soil temperature, and the black line shows
  the baseline of CLM's own differences. The red line shows the difference between CLM-JAX and the unperturbed CLM run.
  Soil variables include all valid soil layers.}
  \label{fig:key8_rmse}
\end{figure}

\begin{figure}[!htbp]
  \centering
  \centerline{\includegraphics[width=1.0\textwidth,height=0.79\textheight,keepaspectratio]{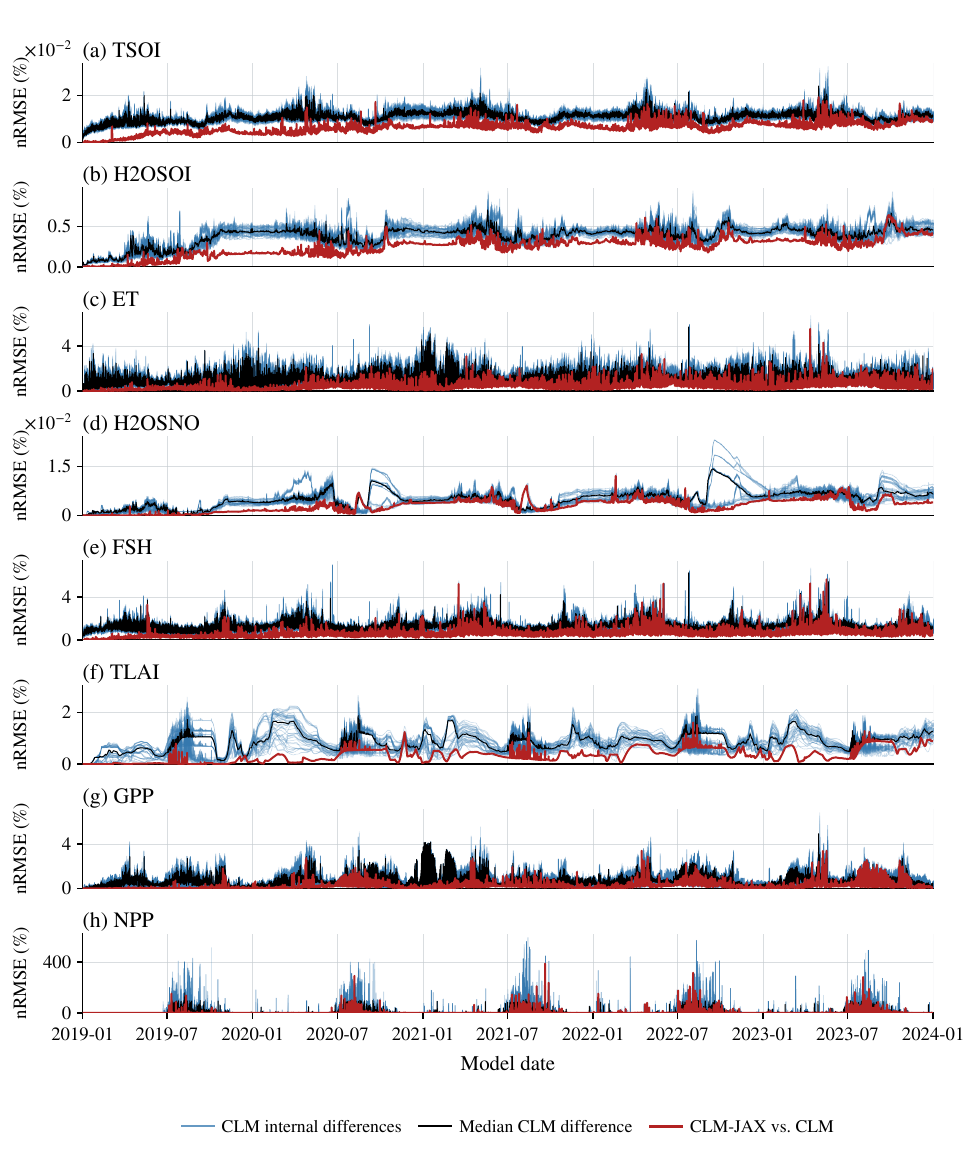}}
  \caption{Six-hourly nRMSE (\%) of eight variables during 2019--2023. Blue lines show the 45
  pairwise nRMSE values among ten CLM runs with perturbed initial soil temperature, and the black
  line shows the baseline of CLM's own differences. The red line shows the nRMSE between CLM-JAX and the unperturbed CLM run.
  For each CLM pair $(i,j)$ with $i<j$, member $i$ is used as the reference.}
  \label{fig:key8_nrmse}
\end{figure}

\begin{figure}[!htbp]
  \centering
  \centerline{\includegraphics[width=0.9\textwidth,height=0.79\textheight,keepaspectratio]{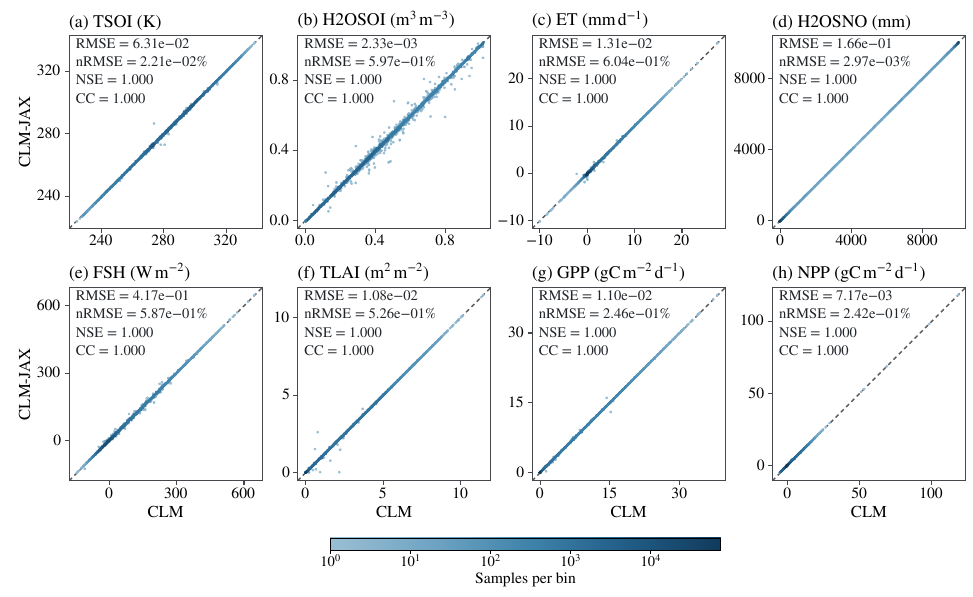}}
  \caption{Six-hourly CLM-JAX versus CLM outputs during 2019--2023. For each variable, 100,000
  paired samples are randomly selected across time and land gridcells. Colors show the number of
  samples in each bin, and dashed lines show $y=x$. Soil variables use the first soil layer.}
  \label{fig:key8_scatter}
\end{figure}

The paired six-hourly samples in Fig.~\ref{fig:key8_scatter} cluster around the 1:1 line for all eight variables. Their nRMSE ranges from 0.00297\% for H2OSNO to 0.604\% for ET, and NSE and CC are close to one. Fig.~S16 in the Supplement shows the paired samples for the remaining variables.

\begin{figure}[!htbp]
  \centering
  \centerline{\includegraphics[width=0.9\textwidth,height=0.79\textheight,keepaspectratio]{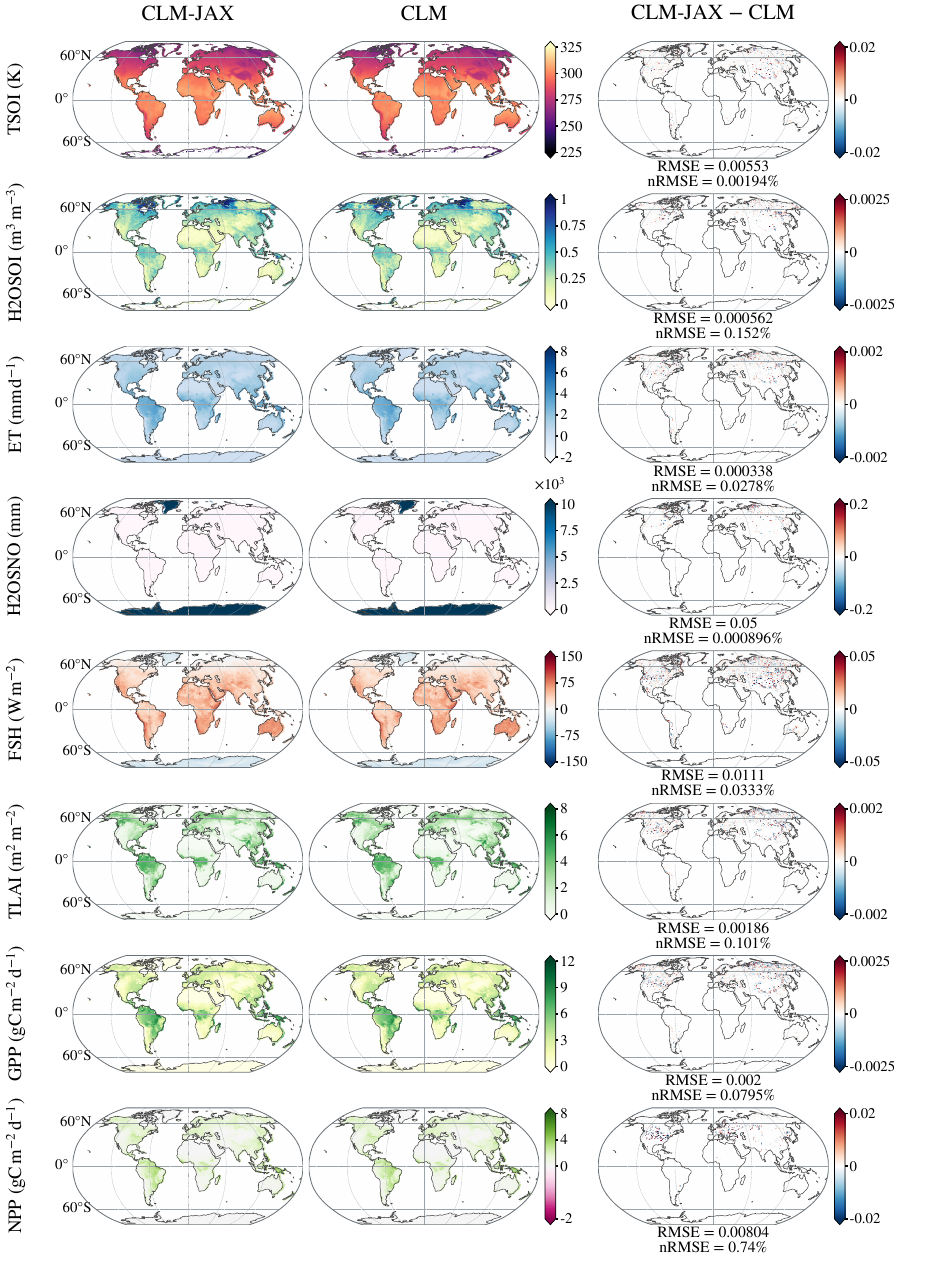}}
  \caption{Mean fields of CLM-JAX, CLM and CLM-JAX minus CLM for eight variables during 2019--2023.
  Soil variables use the first soil layer. RMSE and nRMSE are computed from the mean fields.}
  \label{fig:key8_five_year_meansmotaig}
\end{figure}

The two models produce similar five-year mean spatial patterns in Fig.~\ref{fig:key8_five_year_meansmotaig}. The nRMSE of the mean fields is below 1\% for all eight variables, ranging from 0.000896\% for H2OSNO to 0.740\% for NPP. Figs.~S17--S19 in the Supplement show the mean fields for the remaining variables. Across all 32 variables, the largest mean-field nRMSE is 2.60\% for net ecosystem carbon exchange (NEE), followed by 1.03\% for water intercepted by the canopy (H2OCAN).

\begin{figure}[!htbp]
  \centering
  \centerline{\includegraphics[width=0.9\textwidth,height=0.79\textheight,keepaspectratio]{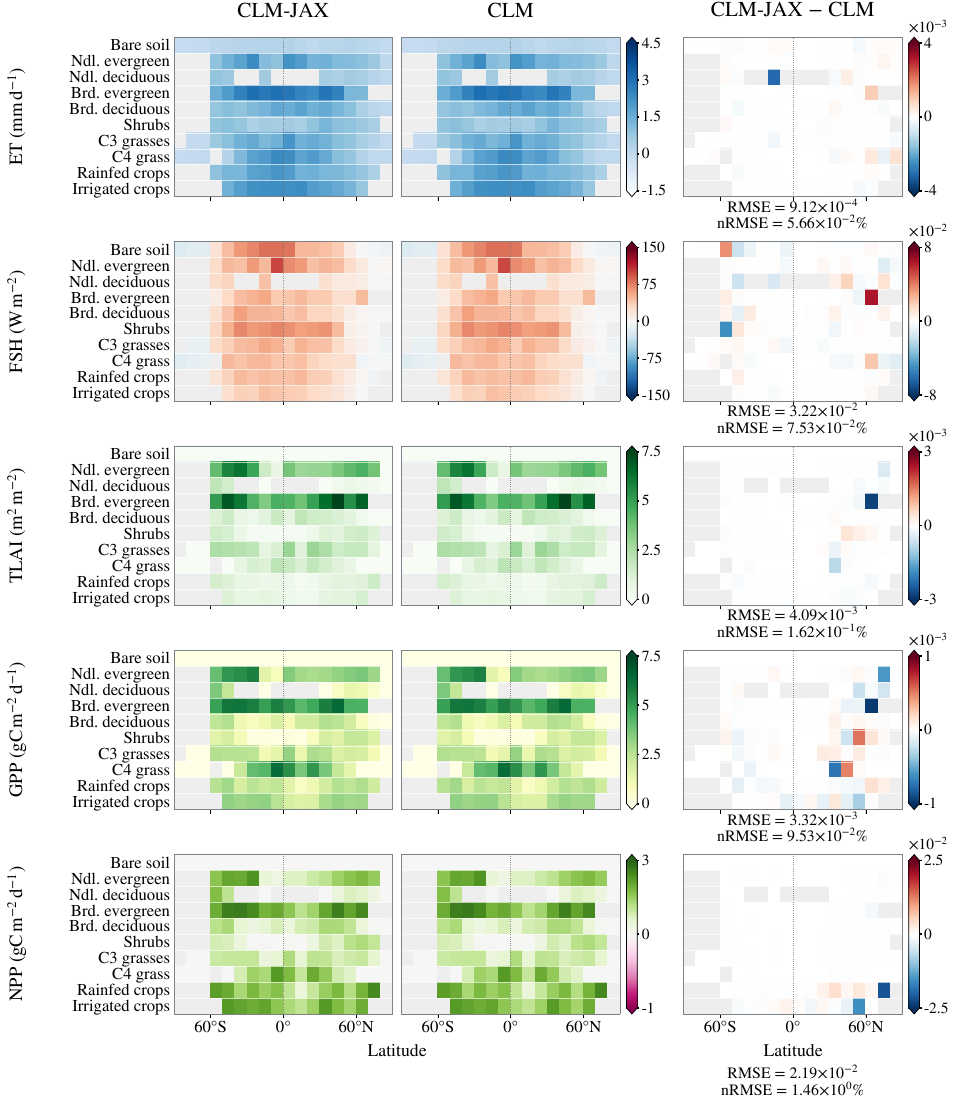}}
  \caption{Area-weighted mean values of CLM-JAX, CLM and CLM-JAX minus CLM by PFT group and
  $10^{\circ}$ latitude band during 2019--2023. Ndl. and Brd. denote needleleaf and broadleaf.
  RMSE and nRMSE are computed from patch-level means.}
  \label{fig:pft_patterns}
\end{figure}

Below the gridcell scale, the two models also agree when five-year means are grouped by PFT and $10^{\circ}$ latitude band. The PFT comparison in Fig.~\ref{fig:pft_patterns} includes ten PFT groups: bare soil, evergreen and deciduous needleleaf trees, evergreen and deciduous broadleaf trees, shrubs, C$_3$ and C$_4$ grasses, and rainfed and irrigated crops. ET, FSH, TLAI, GPP and NPP show similar latitudinal patterns in the two models, with nRMSE below 1.5\% for all five variables. Figs.~S20 and~S21 in the Supplement show additional PFT variables.

Fig.~\ref{fig:column_patterns} compares the five-year means of TSOI, H2OSOI and H2OSNO across eleven CLM column types: natural soil, rainfed and irrigated cropland, lakes, wetlands, glaciers, urban roofs, sunlit and shaded walls, and impervious and pervious roads. Urban density classes are combined within each surface type, and glacier elevation classes are combined. Variables for nonsoil columns follow their CLM output definitions. The nRMSE values for TSOI, H2OSOI and H2OSNO are all below 0.1\%. Figs.~S22 and~S23 in the Supplement extend the comparisons to other water and soil carbon variables.

\begin{figure}[!htbp]
  \centering
  \centerline{\includegraphics[width=0.9\textwidth,height=0.79\textheight,keepaspectratio]{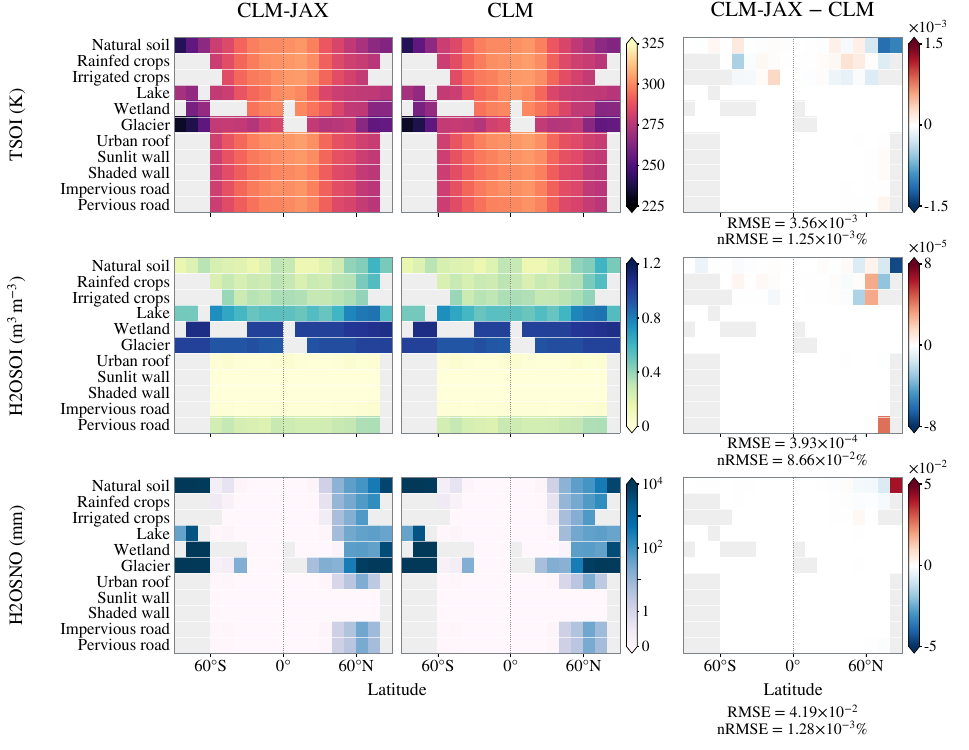}}
  \caption{Area-weighted mean values of CLM-JAX, CLM and CLM-JAX minus CLM by CLM column type and
  $10^{\circ}$ latitude band during 2019--2023. Soil variables use the first layer, centered at
  1\,cm for soil columns. RMSE and nRMSE are computed from column-level means.}
  \label{fig:column_patterns}
\end{figure}

These results indicate that, for the evaluated configuration, the differences between CLM-JAX and CLM remain within the variability that CLM itself produces under round-off-level perturbations.

\subsection{Computational performance}
\label{sec:computational-performance}

Table~\ref{tab:clm-jax-performance} summarizes the sustained integration performance of CLM and CLM-JAX over the five-year integrations. All experiments in this study are performed on a single server with 64 CPU cores and one NVIDIA A800 GPU. On the GPU, CLM-JAX requires 0.7692~s per model hour, with a mean GPU utilization of 96.80\%.

On 64 CPU cores, CLM requires 3.1095 and 1.9805~s per model hour with gfortran and ifort, respectively. Taking the single-core runs as the reference and assuming that the computational speed scales linearly with the number of cores, the speed of CLM-JAX on one GPU would be equivalent to that of CLM on about 64 and 51 cores for gfortran and ifort, respectively. In practice, however, the speed of CLM does not scale linearly with the core count: on 64 cores, CLM is only 15.84 and 19.75 times faster than on one core. As a result, on the same server, CLM-JAX on the GPU is 4.04 and 2.57 times faster than CLM using all 64 CPU cores.

The ifort build is faster at all tested core counts, so the GPU speedup depends on the compiler used for the CPU baseline. Mean CPU utilization, normalized by the core count, remains between 99.82\% and 100.00\% across the tests, indicating that the measurements were not affected by other programs running concurrently on the server.

\begin{table}[!htbp]
  \centering
  \begin{threeparttable}
    \caption{Sustained integration performance of CLM and CLM-JAX in double precision.}
    \label{tab:clm-jax-performance}
    \rmfamily\small
    \setlength{\tabcolsep}{3pt}
    \renewcommand{\arraystretch}{1.3}
    \begin{tabular*}{\textwidth}{@{\extracolsep{\fill}}lllcrrr@{}}
    \toprule
    Model & \shortstack[l]{Computing\\platform} & \shortstack[l]{Compiler or\\framework} & Parallelism & \shortstack{Integration time\\(s per model hour)} & \shortstack{Mean CPU\\utilization (\%)} & \shortstack{Mean GPU\\utilization (\%)} \\
    \midrule
    CLM-JAX & NVIDIA A800 & JAX & 1 GPU & 0.7692 & -- & 96.80 \\
    \midrule
    \multirow{14}{*}{CLM} & \multirow{14}{*}{\shortstack[l]{AMD EPYC\\7542}} & gfortran & 1 core & 49.2395 & 100.00 & -- \\
     &  & ifort & 1 core & 39.1189 & 100.00 & -- \\
     &  & gfortran & 2 cores & 24.8582 & 100.00 & -- \\
     &  & ifort & 2 cores & 21.2230 & 100.00 & -- \\
     &  & gfortran & 4 cores & 13.3168 & 99.98 & -- \\
     &  & ifort & 4 cores & 11.1677 & 100.00 & -- \\
     &  & gfortran & 8 cores & 6.8561 & 99.96 & -- \\
     &  & ifort & 8 cores & 6.4235 & 99.99 & -- \\
     &  & gfortran & 16 cores & 4.1474 & 99.97 & -- \\
     &  & ifort & 16 cores & 3.9741 & 99.97 & -- \\
     &  & gfortran & 32 cores & 3.2071 & 99.96 & -- \\
     &  & ifort & 32 cores & 2.9581 & 99.82 & -- \\
     &  & gfortran & 64 cores & 3.1095 & 99.96 & -- \\
     &  & ifort & 64 cores & 1.9805 & 99.89 & -- \\
    \bottomrule
    \end{tabular*}
    \begin{tablenotes}[flushleft]
      \footnotesize
      \item[] The CPU and GPU tests use the same server. Both models use the \texttt{I2000Clm50BgcCrop} configuration of CESM2.1.3 on the \texttt{f09\_g17} grid, starting from the same initial state on 1 January 2019 and integrating continuously for five years (87,600 time steps). CLM is compiled with GNU Fortran 8.3.0 or Intel ifort 2021.10.0 at optimization level \texttt{-O2}, with other options following the CIME defaults. Integration times exclude initialization, file output and JAX compilation; CLM times are land component times from the CESM timing output. CPU utilization is the summed utilization of all MPI processes divided by the core count. Dashes indicate values that are not applicable.
    \end{tablenotes}
  \end{threeparttable}
\end{table}

\subsection{Gradient verification}
\label{sec:gradient-results}

The gradient test covers 21,013 gridcells, and directional derivatives are evaluated along eight random directions at each gridcell, giving 168,104 gridcell--direction combinations (Table~\ref{tab:gradient-reference-counts}). Of these gridcells, 5,736 have zero soil and crop area and therefore no valid TSOI output, so their 45,888 combinations are excluded from the comparison. The remaining 15,277 gridcells contribute 122,216 combinations. For 87,475 of these, the finite differences satisfy the convergence criterion of Eq.~\eqref{eq:gradient-fd-convergence} and provide reference values; for the other 34,741, the finite differences do not converge. This is due to gradient discontinuities in the numerical model. Perturbations along certain directions can cause processes in the parameterization schemes to switch branches, producing jumps in the gradient and preventing the finite differences from converging. These non-converged perturbation directions therefore do not provide reference values.

\begin{table}[!htbp]
  \centering
  \caption{Availability of finite-difference references in the gradient test. Counts are gridcell--direction combinations. The three categories are mutually exclusive and sum to 168,104. Convergence is determined from forward integrations only, without using the backpropagated gradient.}
  \label{tab:gradient-reference-counts}
  \begin{tabular}{lr}
    \toprule
    Status & Combinations \\
    \midrule
    Converged finite difference & 87,475 \\
    Non-converged finite difference & 34,741 \\
    No TSOI target (zero soil and crop area) & 45,888 \\
    \midrule
    Total & 168,104 \\
    \bottomrule
  \end{tabular}
\end{table}

\begin{figure}[!htbp]
  \centering
  \includegraphics[width=\textwidth,height=0.79\textheight,keepaspectratio]{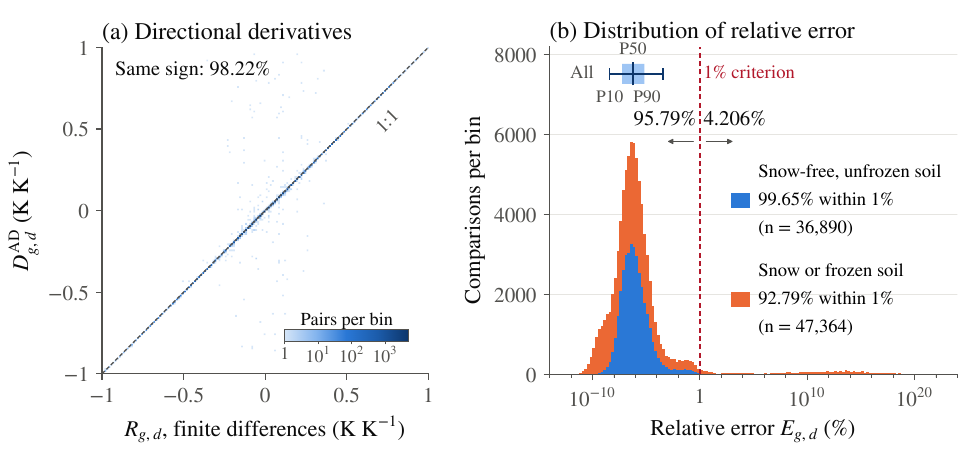}
  \caption{Comparison of the backpropagated gradient with the finite-difference references for all 84,254 nonzero references. (a)~Directional projection $D^{\mathrm{AD}}_{g,d}$ versus reference value $R_{g,d}$, shown only for the range $-1$ to 1~K~K$^{-1}$. Colors indicate the number of combinations in each bin, and the dashed line is the $1\!:\!1$ line. (b)~Histogram of the relative error $E_{g,d}$ with bins 0.25 orders of magnitude wide, stacked by the initial thermal state of the gridcell: snow-free, unfrozen (blue) and snow-covered or frozen (orange). The box plot at the top marks the median (thick vertical line), P25--P75 (box) and P10--P90 (whiskers) of all combinations, and the dashed line marks the 1\% criterion. One combination with a relative error above $10^{24}$\% is included in the rightmost bin.}
  \label{fig:gradient-accuracy}
\end{figure}

The directional projections of the backpropagated gradient and the finite-difference references lie close to the $1\!:\!1$ line (Fig.~\ref{fig:gradient-accuracy}a). The absolute values of the 84,254 nonzero references range from $1.876\times10^{-9}$ to $2.118\times10^{4}$~K~K$^{-1}$, spanning about 13 orders of magnitude, and 82,756 of them (98.22\%) have the same sign as the directional projection of the backpropagated gradient. The median relative error is $5.862\times10^{-7}$\% and the 90th percentile is $3.578\times10^{-4}$\% (Fig.~\ref{fig:gradient-accuracy}b; Table~\ref{tab:gradient-error-percentiles}). Relative errors are at most 1\% for 80,710 references (95.79\%). Of the 3,221 combinations with a zero reference, 3,216 satisfy the absolute error tolerance of Eq.~\eqref{eq:gradient-zero-tolerance}. In total, 83,926 of the 87,475 converged combinations (95.94\%) pass the agreement test, comprising 80,710 nonzero references that meet the 1\% criterion and 3,216 zero references that meet the absolute tolerance.

Snow-free gridcells show markedly higher consistency in the gradient test. The 5,253 snow-free, unfrozen gridcells have 36,890 nonzero references, of which 99.65\% have relative errors of at most 1\%, with a median of $6.487\times10^{-7}$\%. The 10,024 snow-covered or frozen gridcells have 47,364 nonzero references, of which 92.79\% meet the 1\% criterion, with a median of $5.153\times10^{-7}$\%. Finite differences in snow-covered or frozen gridcells also converge less often, with a convergence rate of 62.23\% compared with 89.40\% in snow-free, unfrozen gridcells.

\begin{table}[!htbp]
  \centering
  \caption{Percentiles of the relative error $E_{g,d}$ of the directional projection of the backpropagated gradient. The sample comprises all 84,254 nonzero references, including those with relative errors above 1\%. Combinations with a zero reference are evaluated separately with Eq.~\eqref{eq:gradient-zero-tolerance} and are not included.}
  \label{tab:gradient-error-percentiles}
  \begin{tabular}{lr}
    \toprule
    Percentile & Relative error $E_{g,d}$ \\
    \midrule
    10th & $3.977\times10^{-9}\,\%$ \\
    25th & $5.905\times10^{-8}\,\%$ \\
    50th & $5.862\times10^{-7}\,\%$ \\
    75th & $6.695\times10^{-6}\,\%$ \\
    90th & $3.578\times10^{-4}\,\%$ \\
    \bottomrule
  \end{tabular}
\end{table}

\begin{figure}[!htbp]
  \centering
  \includegraphics[width=\textwidth,height=0.79\textheight,keepaspectratio]{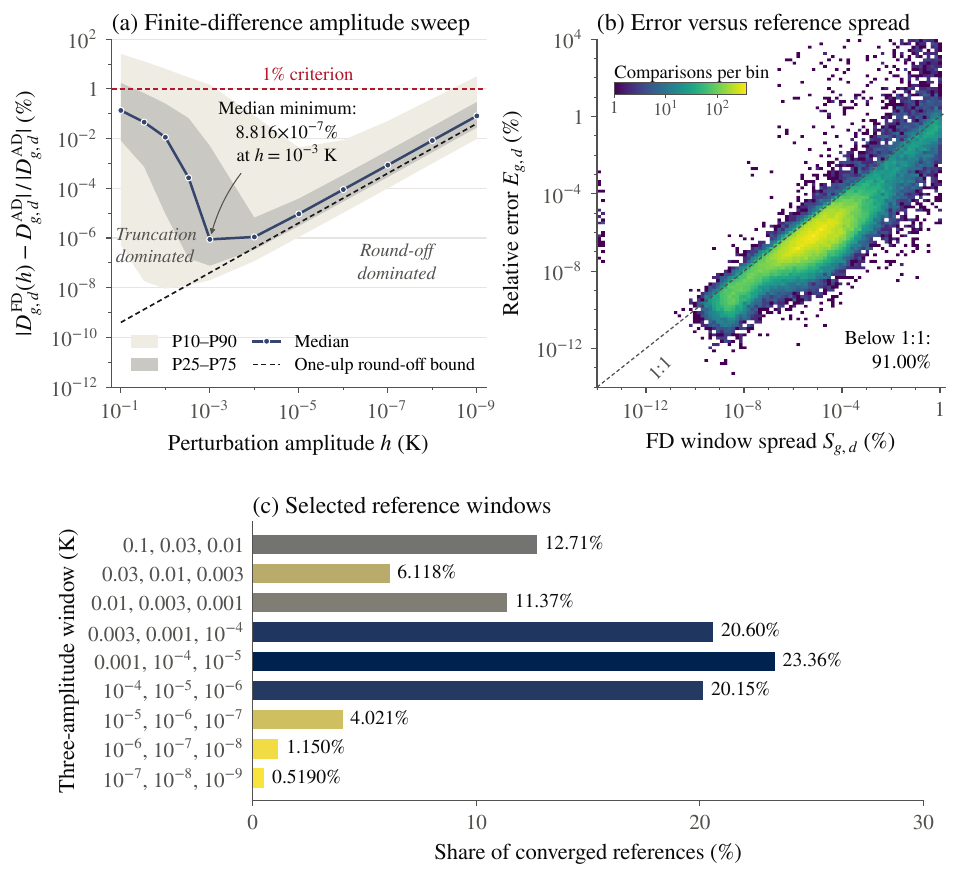}
  \caption{Variation of the finite-difference error with perturbation amplitude and distribution of the selected reference windows. (a)~Deviation $\epsilon_{g,d}(h)$ of the central difference from the directional projection of the backpropagated gradient at each perturbation amplitude (Eq.~\eqref{eq:gradient-amplitude-sweep}) for the 84,254 nonzero references. The solid line shows the median, the shading shows the P25--P75 and P10--P90 ranges, and the dashed line shows the median of the one-ulp round-off bound $\beta_{g,d}(h)$. (b)~Relative error $E_{g,d}$ versus the spread $S_{g,d}$ of the selected window for the 84,145 combinations with $S_{g,d}>0$; by Eq.~\eqref{eq:gradient-fd-convergence}, $S_{g,d}\leq1\%$. Colors indicate the number of combinations in each bin, and the dashed line is the $1\!:\!1$ line. (c)~Share of the 87,475 converged combinations whose reference window is each three-amplitude window. Darker colors indicate larger shares.}
  \label{fig:gradient-fd-convergence}
\end{figure}

To examine how the finite-difference error varies with perturbation amplitude, the central differences at all 11 amplitudes are compared with the directional projection of the backpropagated gradient (Fig.~\ref{fig:gradient-fd-convergence}a). As $h$ decreases, the median relative deviation $\epsilon_{g,d}(h)$ first falls and then rises, forming a V shape. As $h$ decreases from 0.1~K to $10^{-3}$~K, the median falls from 0.1357\% to $8.816\times10^{-7}$\%, and the deviation over this range comes mainly from the truncation error of the central difference. As $h$ decreases further, the median increases in proportion to $1/h$, parallel to the one-ulp round-off bound $\beta_{g,d}(h)$, with a ratio of 2.024--2.740 between the two for $h\leq10^{-4}$~K, indicating that the deviation over this range comes from floating-point round-off. The deviations on the two sides of the V thus follow the behavior expected of truncation and round-off errors, showing that the differences between the two methods arise mainly from the finite differences themselves.

The reference windows selected automatically by minimum spread (Section~\ref{sec:gradient-validation}) lie mostly near the bottom of the V (Fig.~\ref{fig:gradient-fd-convergence}c). Of the 87,475 converged combinations, 64.11\% have reference windows centered at amplitudes between $10^{-5}$ and $10^{-3}$~K, 30.20\% at larger amplitudes and 5.690\% at smaller amplitudes. As a result, the median relative error in Table~\ref{tab:gradient-error-percentiles} ($5.862\times10^{-7}$\%) is of the same order as the median deviation at the bottom of the V ($8.816\times10^{-7}$\%). The spread $S_{g,d}$ of the three finite-difference estimates in the selected window reflects the uncertainty of the reference itself. Of the 84,145 combinations with $S_{g,d}>0$, 91.00\% have a relative error $E_{g,d}$ no larger than $S_{g,d}$ (Fig.~\ref{fig:gradient-fd-convergence}b); that is, the difference between the directional projection of the backpropagated gradient and the reference does not exceed the uncertainty of the reference itself.

\section{Discussion}
\label{sec:discussion}

\subsection{Main contributions}
\label{sec:discussion-context}

The main contribution of this work is that CLM-JAX reimplements the process equations, subgrid hierarchy and process coupling of CLM in a widely used scientific configuration, while supporting GPU acceleration and end-to-end differentiability. Data-driven AI models can reduce computational cost, but they operate as black boxes, lack physical interpretability and have difficulty guaranteeing physical consistency \citep{reichstein2019deep}. OpenACC-based GPU ports retain the original Fortran code and can improve computational performance, but OpenACC itself does not provide automatic differentiation \citep{wang2014openacc,schwartz2022openacc}.

In the five-year global integration, 32 variables are evaluated, and CLM-JAX performs well in both the numerical consistency and gradient tests. Such a long integration also verifies the correct implementation of annual phenological updates, which short simulations can easily miss. Previous differentiable implementations of land surface models were mostly evaluated at individual sites or a small number of gridcells \citep{lahlou2026systematic,tian2026noahpy}; they did not establish a numerical consistency baseline, and their gradient error assessments did not reach the $10^{-6}$ level achieved here. The present evaluation covers diverse climates and land surface types while retaining the coupling among energy, water, carbon, nitrogen and crop processes, and the gradient test covers 15,277 gridcells worldwide, exceeding the scale of previous tests at individual sites or a few gridcells.

\subsection{Numerical consistency and Computational performance}
\label{sec:discussion-consistency}

We establish a stringent baseline of CLM's own differences to test the numerical correctness of the CLM-JAX implementation. This approach draws on the ensemble consistency test described by \citet{baker2015pycect} but differs from the original method. Perturbations of order $10^{-14}$~K are added to the initial soil temperature of ten CLM members, and the differences among the members serve as the baseline of CLM's own differences, against which the differences between CLM-JAX and CLM are measured. These differences arise entirely from floating-point perturbations of 1~ulp in the initial state. The errors between CLM-JAX and CLM remain almost always below this baseline; even during periods when the RMSE between CLM-JAX and CLM is relatively large, it remains below the baseline of CLM's own differences. This indicates that the scientific implementations of CLM-JAX and CLM can be regarded as identical, and that the remaining differences between them can be attributed entirely to floating-point errors.

With CLM-JAX, a single NVIDIA A800 achieves sustained integration speeds 4.04 and 2.57 times those of CLM compiled with gfortran and ifort, respectively, on 64 CPU cores (Table~\ref{tab:clm-jax-performance}). Land surface calculations offer substantial parallelism across columns and patches, while vertical recurrences within each column retain their original order. CLM-JAX combines related array operations within kernels, reuses compiled functions and keeps static data on the device, reducing intermediate data transfers and repeated setup costs.

\subsection{Differentiability}
\label{sec:discussion-differentiability}

The results in Section~\ref{sec:gradient-results} show that the backpropagated gradients of CLM-JAX agree closely with the finite-difference results. Of the 87,475 gridcell--direction combinations for which the finite differences converge and provide a reference value, 95.94\% satisfy the agreement criterion, that is, a relative error of at most 1\% for nonzero references or the absolute error tolerance for zero references. For nonzero references, the median relative error is $5.862\times10^{-7}$\%, and the derivative signs agree in 98.22\% of cases. The remaining differences arise mainly from errors in the finite differences themselves: the median relative error is of the same order as the finite-difference error at the optimal perturbation amplitude; at large amplitudes the deviation follows the behavior of truncation error, and at small amplitudes it parallels the round-off error bound; and for 91.00\% of gridcell--direction combinations, the difference between the backpropagated gradient and the reference does not exceed the uncertainty of the reference itself. These results show that, within the tested scope, the custom derivative rules and gradient backpropagation through the 100-step coupled integration yield accurate gradients.

The finite differences fail to converge for a considerable fraction of gridcell--direction combinations, which is expected. The model contains many parameterization schemes, and switching between their branches introduces discontinuities in the gradient. We therefore evaluate the gradients using the perturbation directions for which the finite differences converge; these directions involve fewer switches between physical branches and correspond more closely to the gradient propagation path. Soil freezing and thawing provide an example. The phase-change scheme of CLM has two branches separated by the freezing point (273.15~K). When the temperature of a soil layer does not cross the freezing point, no phase change occurs, and perturbations of the initial temperature propagate to the surface layer through heat conduction, giving a finite gradient. When the layer temperature crosses the freezing point and the layer contains ice or liquid water available for phase change, the model enters the phase-change branch: the layer temperature is first reset to the freezing point, and the excess or deficit energy is used for melting or freezing. As long as this energy does not exceed that required for the phase change, the layer temperature remains at the freezing point, the temperature perturbation is entirely converted into changes in ice and liquid water content, and the gradient of the layer temperature with respect to the initial temperature is zero. Because the two branches have different gradients, the gradient of the objective function is discontinuous at the freezing point. For gridcells near the freezing point, finite perturbations cross this boundary, so the central differences at different amplitudes mix the responses of the two branches and vary with amplitude instead of converging. The finite-difference results clearly reflect this property: the convergence rate of the finite differences is 62.23\% in snow-covered or frozen gridcells, markedly lower than the 89.40\% in snow-free, unfrozen gridcells, and this is why the finite differences fail to converge in many directions. Such gradient discontinuities have also been one of the long-standing difficulties in the development of adjoint models.

\subsection{Extensions and future work}
\label{sec:discussion-future}

The I2000Clm50BgcCrop configuration used here is a typical CLM5.0 configuration that includes fully coupled biogeophysical, biogeochemical and crop processes. CLM-JAX uses the CESM formats for input, parameter and restart files and retains interfaces for exchanging boundary conditions with external components, such as the MOSART river exchanges currently supplied as external boundary inputs. It can therefore be extended to other configurations and to scenarios coupled with atmospheric, river and other components.

In future work, we will verify gradients over longer integration periods, replace threshold-type physical parameterizations with AI parameterizations to smooth the gradients, perform end-to-end parameter optimization and four-dimensional variational data assimilation using observations, and train the networks end to end after the AI parameterizations are introduced. These applications will further demonstrate the practical value of CLM-JAX gradients. For example, observational data such as flux-site measurements and remote sensing products are currently used mainly to evaluate land surface models, and their use in initial-state assimilation and parameter optimization remains limited. An important reason is that conventional land surface models lack gradient information, making it difficult to exploit large volumes of observations for efficient optimization. The differentiability of CLM-JAX provides a basis for using these observations in gradient-based applications, such as end-to-end parameter optimization and four-dimensional variational data assimilation \citep{shen2023differentiable}. In the future, CLM-JAX will adopt a hybrid physics--data modeling framework that represents uncertain empirical relationships in the existing process equations while maintaining physical consistency, including the conservation of water, energy and elements.

\conclusions
\label{sec:conclusions}

We develop CLM-JAX, a land surface model implemented in Python and JAX based on the CLM5.0 BGC-Crop configuration in CESM2.1.3. CLM-JAX fully retains the subgrid hierarchy and parameterizations of the original model and the coupling among energy, water, carbon, nitrogen and crop processes, while supporting computation on CPUs and GPUs and end-to-end differentiability.

In five-year continuous global simulations over 2019--2023, 32 variables are evaluated against the stringent baseline of CLM's own numerical differences. The errors between CLM-JAX and CLM remain almost always below this baseline, including during periods of relatively large RMSE, and the maximum nRMSE of the five-year mean fields across the 32 variables is 2.60\%, for net ecosystem carbon exchange. The scientific implementations of CLM-JAX and CLM can therefore be regarded as identical, and the remaining differences between them can be attributed entirely to floating-point errors. In double precision, a single NVIDIA A800 achieves sustained integration speeds 4.04 and 2.57 times those of CLM compiled with gfortran and ifort, respectively, on 64 CPU cores.

The gradient test compares backpropagated gradients with finite-difference results along eight random directions at 15,277 gridcells worldwide. Of the 87,475 gridcell--direction combinations for which the finite differences converge, 95.94\% satisfy the agreement criterion; for nonzero references, the median relative error is $5.862\times10^{-7}$\%, and the derivative signs agree in 98.22\% of cases. The remaining differences arise mainly from the truncation and round-off errors of the finite differences themselves. Combinations without converged finite differences occur mainly in snow-covered or frozen gridcells, reflecting gradient discontinuities caused by branch switching in parameterizations such as soil freezing and thawing.

In the development of CLM-JAX, differences in the handling of some annually updated accumulators appear only after one year of integration. This shows that multiyear comparisons against the baseline of CLM's own differences are a necessary complement to module tests and short simulations, and can inform the reimplementation and verification of other Earth system model components. CLM-JAX uses the CESM file formats and coupling interfaces and can be extended to other configurations and coupled scenarios. Future work will verify gradients over longer integration periods and use observational data, such as flux-site measurements and remote sensing products, for end-to-end parameter optimization and four-dimensional variational data assimilation.

\codeavailability{
The CLM-JAX source code is distributed under the GNU Lesser General Public License version 2.1 (LGPL-2.1). The code repository is \url{https://github.com/zhenw2023/CLM-JAX}, and the Zenodo DOI for the code archive is \url{https://doi.org/10.5281/zenodo.24488761}.
}

\dataavailability{
The WFDE5 v3.0 atmospheric forcing used in this study is available from the Copernicus Climate Change Service Climate Data Store at \url{https://doi.org/10.24381/cds.20d54e34} \citep{c3s2026wfde5}.
}

\authorcontribution{
S.Y.: methodology, writing---review and editing, and funding acquisition;
Z.W.: methodology, software, numerical experiments, data analysis, visualization, writing---original draft, and writing---review and editing;
T.S.: methodology, software, numerical experiments, data analysis, and visualization;
B.M.: supervision and funding acquisition;
B.Q.: methodology and funding acquisition;
G.D.: study design and software.
}

\competinginterests{The authors declare that they have no conflict of interest.}

\begin{acknowledgements}
This study was supported in part by the Meteorological Joint Funds of the National Natural Science Foundation of China
under Grant U2542211, U2542212, and in part by the Original Exploration Project of the National Natural Science Foundation of China
under Grant 42450163, and in part by the National Natural Science Foundation of China under Grant 42635009, 42405147.
\end{acknowledgements}

\bibliographystyle{copernicus}
\bibliography{references}

\end{document}

\title{Supplement of CLM-JAX v3.2.46: a differentiable, GPU-accelerated land surface model numerically indistinguishable from the Community Land Model 5.0}
\Author[1]{Shijin}{Yuan}
\Author[1]{Zhen}{Wei}
\Author[1]{Tianyu}{Shen}
\Author[1][binmu@tongji.edu.cn]{Bin}{Mu}
\Author[1]{Bo}{Qin}
\Author[2]{Guokun}{Dai}
\affil[1]{School of Computer Science and Technology, Tongji University, Shanghai 201804, China}
\affil[2]{Department of Atmospheric and Oceanic Sciences \& Institute of Atmospheric Sciences, Fudan University, Shanghai 200438, China}
\equalcontrib{1,2}

\runningtitle{CLM-JAX}
\runningauthor{Shijin Yuan et al.}
\firstpage{1}

\maketitle
\nolinenumbers
\linespread{1}\selectfont

\section*{Introduction}

This Supplement describes the CLM subgrid structure and presents additional numerical consistency results from the 2019--2023 global simulations. Fig.~\ref{fig:CLM} shows the gridcell, landunit, column and patch hierarchy. Table~\ref{tab:clm5-processes-en} lists the main processes and subprocesses in CLM, and Table~\ref{tab:validation-variables} defines the 32 evaluated variables.

Figs.~\ref{fig:key8_nse} and~\ref{fig:key8_cc} show Nash--Sutcliffe efficiency (NSE) and Pearson correlation coefficient (CC) for the eight variables presented in the main text. Figs.~\ref{fig:state_rmse}--\ref{fig:carbon_cc} report root mean square error (RMSE), normalized RMSE (nRMSE), NSE and CC for the remaining 24 variables. Each time-series plot compares the numerical differences between CLM-JAX and the unperturbed CLM run with the baseline of CLM's own numerical differences.

Fig.~\ref{fig:supp24_scatter} compares paired six-hourly samples for the additional 24 variables, and Figs.~\ref{fig:state_five_year_means}--\ref{fig:carbon_five_year_means} show their five-year mean fields. Figs.~\ref{fig:pft_water}--\ref{fig:column_drainage_carbon} show area-weighted five-year means of additional water, energy and carbon variables by plant functional type group and CLM column type in $10^{\circ}$ latitude bands.

\begin{figure}[!htbp]
  \centering
  \centerline{\includegraphics[width=0.75\textwidth,height=0.83\textheight,keepaspectratio]{smfigure/Sfig01.pdf}}
  \caption{Schematic of the CLM5.0 subgrid hierarchy, comprising gridcells, landunits, columns, and patches. Multiple plant functional types (PFTs) share a soil column within the natural vegetation land unit, with the dashed box highlighting their patch-level representation. Urban columns represent roofs, sunlit and shaded walls, and pervious and impervious canyon floors. Crop columns distinguish rainfed and irrigated conditions for each crop functional type. Lake and glacier substructures are omitted for clarity. Irrig denotes irrigated conditions; N and M denote the numbers of vegetation and crop functional types illustrated schematically.}
  \label{fig:CLM}
\end{figure}

\begin{figure}
  \centering
  \centerline{\includegraphics[width=1.0\textwidth,height=0.83\textheight,keepaspectratio]{smfigure/Sfig02.pdf}}
  \caption{Six-hourly Nash--Sutcliffe efficiency (NSE) of eight variables during 2019--2023. Blue lines show
  the 45 pairwise NSE values among ten CLM runs with perturbed initial soil temperature, and the
  black line shows the baseline of CLM's own differences. The red line shows the NSE between CLM-JAX and the reference CLM
  run. Soil variables include all valid soil layers.}
  \label{fig:key8_nse}
\end{figure}

\begin{figure}
  \centering
  \centerline{\includegraphics[width=1.0\textwidth,height=0.83\textheight,keepaspectratio]{smfigure/Sfig03.pdf}}
  \caption{Six-hourly Pearson correlation coefficient (CC) of the eight variables in Fig.~\ref{fig:key8_nse}
  during 2019--2023. Comparisons and line colors follow Fig.~\ref{fig:key8_nse}. Soil variables
  include all valid soil layers.}
  \label{fig:key8_cc}
\end{figure}

\clearpage

\begin{figure}[p]
  \centering
  \includegraphics[width=1.0\textwidth,height=0.83\textheight,keepaspectratio]{smfigure/Sfig04.pdf}
  \caption{Six-hourly RMSE of eight additional state variables during 2019--2023. Blue lines show the 45
  pairwise differences among ten CLM runs with perturbed initial soil temperature, and the black
  line shows the baseline of CLM's own differences. The red line shows the difference between CLM-JAX and the reference CLM
  run. Soil variables include all valid soil layers.}
  \label{fig:state_rmse}
\end{figure}

\clearpage

\begin{figure}[p]
  \centering
  \includegraphics[width=1.0\textwidth,height=0.83\textheight,keepaspectratio]{smfigure/Sfig05.pdf}
  \caption{Six-hourly RMSE of eight additional water and energy fluxes during 2019--2023. Line colors follow
  Fig.~\ref{fig:state_rmse}.}
  \label{fig:flux_rmse}
\end{figure}

\clearpage

\begin{figure}[p]
  \centering
  \includegraphics[width=1.0\textwidth,height=0.83\textheight,keepaspectratio]{smfigure/Sfig06.pdf}
  \caption{Six-hourly RMSE of eight additional energy and carbon variables during 2019--2023. Line colors
  follow Fig.~\ref{fig:state_rmse}.}
  \label{fig:carbon_rmse}
\end{figure}

\clearpage

\begin{figure}[p]
  \centering
  \includegraphics[width=1.0\textwidth,height=0.83\textheight,keepaspectratio]{smfigure/Sfig07.pdf}
  \caption{Six-hourly nRMSE (\%) of the state variables in Fig.~\ref{fig:state_rmse} during 2019--2023. Blue
  lines show the 45 pairwise nRMSE values among ten CLM runs with perturbed initial soil
  temperature, and the black line shows the baseline of CLM's own differences. The red line shows the nRMSE between CLM-JAX
  and the reference CLM run. For each CLM pair $(i,j)$ with $i<j$, member $i$ is used as the
  reference.}
  \label{fig:state_nrmse}
\end{figure}

\clearpage

\begin{figure}[p]
  \centering
  \includegraphics[width=1.0\textwidth,height=0.83\textheight,keepaspectratio]{smfigure/Sfig08.pdf}
  \caption{Six-hourly nRMSE (\%) of the water and energy fluxes in Fig.~\ref{fig:flux_rmse} during 2019--2023.
  Line colors follow Fig.~\ref{fig:state_nrmse}.}
  \label{fig:flux_nrmse}
\end{figure}

\clearpage

\begin{figure}[p]
  \centering
  \includegraphics[width=1.0\textwidth,height=0.83\textheight,keepaspectratio]{smfigure/Sfig09.pdf}
  \caption{Six-hourly nRMSE (\%) of the energy and carbon variables in Fig.~\ref{fig:carbon_rmse} during
  2019--2023. Line colors follow Fig.~\ref{fig:state_nrmse}.}
  \label{fig:carbon_nrmse}
\end{figure}

\clearpage

\begin{figure}[p]
  \centering
  \includegraphics[width=1.0\textwidth,height=0.83\textheight,keepaspectratio]{smfigure/Sfig10.pdf}
  \caption{Six-hourly NSE of the state variables in Fig.~\ref{fig:state_rmse} during 2019--2023. Line colors
  follow Fig.~\ref{fig:key8_nse}. Soil variables include all valid soil layers.}
  \label{fig:state_nse}
\end{figure}

\clearpage

\begin{figure}[p]
  \centering
  \includegraphics[width=1.0\textwidth,height=0.83\textheight,keepaspectratio]{smfigure/Sfig11.pdf}
  \caption{Six-hourly NSE of the water and energy fluxes in Fig.~\ref{fig:flux_rmse} during 2019--2023. Line
  colors follow Fig.~\ref{fig:key8_nse}.}
  \label{fig:flux_nse}
\end{figure}

\clearpage

\begin{figure}[p]
  \centering
  \includegraphics[width=1.0\textwidth,height=0.83\textheight,keepaspectratio]{smfigure/Sfig12.pdf}
  \caption{Six-hourly NSE of the energy and carbon variables in Fig.~\ref{fig:carbon_rmse} during 2019--2023.
  Line colors follow Fig.~\ref{fig:key8_nse}.}
  \label{fig:carbon_nse}
\end{figure}

\clearpage

\begin{figure}[p]
  \centering
  \includegraphics[width=1.0\textwidth,height=0.83\textheight,keepaspectratio]{smfigure/Sfig13.pdf}
  \caption{Six-hourly CC of the state variables in Fig.~\ref{fig:state_rmse} during 2019--2023. Line colors
  follow Fig.~\ref{fig:key8_cc}. Soil variables include all valid soil layers.}
  \label{fig:state_cc}
\end{figure}

\clearpage

\begin{figure}[p]
  \centering
  \includegraphics[width=1.0\textwidth,height=0.83\textheight,keepaspectratio]{smfigure/Sfig14.pdf}
  \caption{Six-hourly CC of the water and energy fluxes in Fig.~\ref{fig:flux_rmse} during 2019--2023. Line
  colors follow Fig.~\ref{fig:key8_cc}.}
  \label{fig:flux_cc}
\end{figure}

\clearpage

\begin{figure}[p]
  \centering
  \includegraphics[width=1.0\textwidth,height=0.83\textheight,keepaspectratio]{smfigure/Sfig15.pdf}
  \caption{Six-hourly CC of the energy and carbon variables in Fig.~\ref{fig:carbon_rmse} during 2019--2023.
  Line colors follow Fig.~\ref{fig:key8_cc}.}
  \label{fig:carbon_cc}
\end{figure}

\clearpage

\begin{figure}[p]
  \centering
  \includegraphics[width=0.9\textwidth,height=0.83\textheight,keepaspectratio]{smfigure/Sfig16.pdf}
  \caption{Six-hourly CLM-JAX versus CLM outputs for 24 additional variables during 2019--2023. Panels a--h
  show state variables, i--p show water and energy fluxes, and q--x show energy and carbon
  variables. Colors show the number of samples in each bin, and dashed lines show $y=x$. SOILLIQ and
  SOILICE use the first soil layer.}
  \label{fig:supp24_scatter}
\end{figure}

\clearpage

\begin{figure}[p]
  \centering
  \includegraphics[width=0.9\textwidth,height=0.83\textheight,keepaspectratio]{smfigure/Sfig17.pdf}
  \caption{Mean fields of the state variables during 2019--2023. Columns show CLM-JAX, CLM and CLM-JAX minus
  CLM. SOILLIQ and SOILICE use the first soil layer. RMSE and nRMSE are computed from the
  mean fields.}
  \label{fig:state_five_year_means}
\end{figure}

\clearpage

\begin{figure}[p]
  \centering
  \includegraphics[width=0.9\textwidth,height=0.83\textheight,keepaspectratio]{smfigure/Sfig18.pdf}
  \caption{Mean fields of the water and energy fluxes during 2019--2023. Columns show CLM-JAX, CLM and CLM-JAX
  minus CLM. RMSE and nRMSE are computed from the mean fields.}
  \label{fig:flux_five_year_means}
\end{figure}

\clearpage

\begin{figure}[p]
  \centering
  \includegraphics[width=0.9\textwidth,height=0.83\textheight,keepaspectratio]{smfigure/Sfig19.pdf}
  \caption{Mean fields of the energy and carbon variables during 2019--2023. Columns show CLM-JAX, CLM and
  CLM-JAX minus CLM. RMSE and nRMSE are computed from the mean fields.}
  \label{fig:carbon_five_year_means}
\end{figure}

\clearpage

\begin{figure}[p]
  \centering
  \includegraphics[width=0.9\textwidth,height=0.83\textheight,keepaspectratio]{smfigure/Sfig20.pdf}
  \caption{Area-weighted means of water storage and evaporative fluxes by PFT group and $10^{\circ}$ latitude
  band during 2019--2023. Columns show CLM-JAX, CLM and CLM-JAX minus CLM. Ndl. and Brd. denote
  needleleaf and broadleaf. RMSE and nRMSE are computed from patch-level means.}
  \label{fig:pft_water}
\end{figure}

\clearpage

\begin{figure}[p]
  \centering
  \includegraphics[width=0.9\textwidth,height=0.83\textheight,keepaspectratio]{smfigure/Sfig21.pdf}
  \caption{Area-weighted means of energy and carbon variables by PFT group and $10^{\circ}$ latitude band
  during 2019--2023. Columns show CLM-JAX, CLM and CLM-JAX minus CLM. Ndl. and Brd. denote
  needleleaf and broadleaf. RMSE and nRMSE are computed from patch-level means.}
  \label{fig:pft_energy_carbon}
\end{figure}

\clearpage

\begin{figure}[p]
  \centering
  \includegraphics[width=0.9\textwidth,height=0.83\textheight,keepaspectratio]{smfigure/Sfig22.pdf}
  \caption{Area-weighted means of water storage and total runoff by CLM column type and $10^{\circ}$ latitude
  band during 2019--2023. Columns show CLM-JAX, CLM and CLM-JAX minus CLM. SOILLIQ and SOILICE use
  the first layer, centered at 1\,cm for soil columns. RMSE and nRMSE are computed from column-level
  means.}
  \label{fig:column_water}
\end{figure}

\clearpage

\begin{figure}[p]
  \centering
  \includegraphics[width=0.9\textwidth,height=0.83\textheight,keepaspectratio]{smfigure/Sfig23.pdf}
  \caption{Area-weighted means of drainage and soil carbon by CLM column type and $10^{\circ}$ latitude band
  during 2019--2023. Columns show CLM-JAX, CLM and CLM-JAX minus CLM. RMSE and nRMSE are computed
  from column-level means.}
  \label{fig:column_drainage_carbon}
\end{figure}

\clearpage

\begin{table}[!htbp]
  \centering
  \caption{Major processes and subprocesses in CLM5.0.}
  \label{tab:clm5-processes-en}
  \small
  \setlength{\tabcolsep}{5pt}
  \renewcommand{\arraystretch}{1.3}
  \begin{tabularx}{\textwidth}{@{}c >{\raggedright\arraybackslash}p{0.24\textwidth} >{\raggedright\arraybackslash}X@{}}
  \toprule
  No. & Major process & Subprocesses \\
  \midrule
  1 & Forcing and state preparation & Interpolate and downscale meteorological forcing; update daylength and surface states; apply previously determined irrigation. \\
  2 & Canopy hydrology & Rain and snow interception, canopy water storage, throughfall and canopy drip; update ground precipitation inputs and canopy wetness. \\
  3 & Radiative transfer & Distinguish sunlit and shaded leaves; calculate shortwave absorption and partitioning across vegetation, the ground surface and urban land units. \\
  4 & Surface energy exchange and plant physiology & Solve bare-ground and canopy energy balances; couple C$_3$/C$_4$ photosynthesis, stomatal conductance and plant hydraulics; update leaf nitrogen allocation and photosynthetic capacity; calculate urban and lake surface fluxes and subsequent irrigation demand. \\
  5 & Soil, snow and lake thermal processes & Compute heat transfer and freezing--thawing in lakes and soil--snow columns; update temperature, ice--water partitioning and surface heat fluxes. \\
  6 & Land surface water transport & Snowmelt percolation, surface runoff, infiltration, soil water redistribution and root water uptake; adjust snow layers, calculate glacier surface mass balance, and update lake water and snow grain properties. \\
  7 & Carbon--nitrogen cycling and crop growth & Nitrogen deposition, fixation and fertilizer inputs; plant maintenance respiration.\newline Potential decomposition, nitrification and denitrification.\newline Natural vegetation phenology; crop planting, growth, grain filling and maturity; nitrogen acquisition and its carbon costs, nutrient competition and carbon--nitrogen allocation.\newline Actual decomposition, mineralization and immobilization; tissue turnover, litterfall and transfer of harvested biomass to product pools; growth respiration.\newline Carbon and nitrogen pool updates and vertical transport; vegetation mortality and fire. \\
  8 & Soil drainage and nitrogen leaching & Calculate lateral soil drainage and subsurface runoff; update nitrogen leaching, summarize carbon and nitrogen states, and update canopy structure, including leaf area; check water, energy, carbon and nitrogen balances. \\
  9 & Methane cycling & Methane production and oxidation, diffusion, ebullition and plant-mediated transport; calculate land--atmosphere methane fluxes. \\
  10 & Albedo update & Update surface albedo from the revised vegetation, soil and snow states for radiation calculations in the next time step. \\
  11 & Flux aggregation and output & Aggregate land states and fluxes by area, exchange fields with other components, and write history and restart files at the prescribed output frequency. \\
  \bottomrule
  \end{tabularx}
  \par\smallskip
  \begin{minipage}{\textwidth}\footnotesize
  Note: Processes are grouped in the order of the main calls within a time step. Internal iterations and staged updates are not listed separately.
  \end{minipage}
  \end{table}

\clearpage

\begin{table}[p]
  \centering
  \begin{threeparttable}
    \caption{Variables used to evaluate the numerical consistency of CLM-JAX.}
    \label{tab:validation-variables}
    \rmfamily\small
    \setlength{\tabcolsep}{5pt}
    \renewcommand{\arraystretch}{1.15}
    \begin{tabularx}{\textwidth}{@{}>{\raggedright\arraybackslash}p{0.17\textwidth} >{\raggedright\arraybackslash}p{0.15\textwidth} >{\raggedright\arraybackslash}X >{\raggedright\arraybackslash}p{0.17\textwidth}@{}}
    \toprule
    Category & Variable & Definition & Unit \\
    \midrule
    \multirow[t]{11}{=}{Land temperature and water states} & \textbf{TSOI} & Soil temperature & K \\
     & \textbf{H2OSOI} & Volumetric soil water content & m$^{3}$ m$^{-3}$ \\
     & \textbf{H2OSNO} & Snow water equivalent & mm \\
     & TSA & Air temperature at 2 m & K \\
     & TG & Ground surface temperature & K \\
     & SOILLIQ & Liquid water in each soil layer & kg m$^{-2}$ \\
     & SOILICE & Ice in each soil layer & kg m$^{-2}$ \\
     & H2OCAN & Water intercepted by the canopy & mm \\
     & H2OSFC & Depth of surface water & mm \\
     & TWS & Total terrestrial water storage & mm \\
     & ZWT & Depth to the water table & m \\
    \midrule
    \multirow[t]{7}{=}{Water fluxes} & \textbf{ET} & Total evapotranspiration & mm d$^{-1}$ \\
     & QSOIL & Net ground evaporation & mm d$^{-1}$ \\
     & QVEGE & Evaporation from the canopy & mm d$^{-1}$ \\
     & QVEGT & Plant transpiration & mm d$^{-1}$ \\
     & QRUNOFF & Total liquid runoff & mm d$^{-1}$ \\
     & QOVER & Surface runoff & mm d$^{-1}$ \\
     & QDRAI & Subsurface drainage & mm d$^{-1}$ \\
    \midrule
    \multirow[t]{5}{=}{Energy fluxes} & \textbf{FSH} & Sensible heat flux & W m$^{-2}$ \\
     & LE & Total latent heat flux & W m$^{-2}$ \\
     & FSA & Absorbed shortwave radiation & W m$^{-2}$ \\
     & FIRA & Net longwave radiation & W m$^{-2}$ \\
     & FGR & Heat flux into the ground & W m$^{-2}$ \\
    \midrule
    \multirow[t]{9}{=}{Vegetation and carbon} & \textbf{TLAI} & Total projected leaf area index & m$^{2}$ m$^{-2}$ \\
     & \textbf{GPP} & Gross primary production & g C m$^{-2}$ d$^{-1}$ \\
     & \textbf{NPP} & Net primary production & g C m$^{-2}$ d$^{-1}$ \\
     & AR & Autotrophic respiration & g C m$^{-2}$ d$^{-1}$ \\
     & HR & Heterotrophic respiration & g C m$^{-2}$ d$^{-1}$ \\
     & NEE & Net ecosystem carbon exchange & g C m$^{-2}$ d$^{-1}$ \\
     & TOTVEGC & Vegetation carbon, excluding the temporary C pool & g C m$^{-2}$ \\
     & TOTLITC & Carbon stored in litter & g C m$^{-2}$ \\
     & TOTSOMC & Carbon in soil organic matter & g C m$^{-2}$ \\
    \bottomrule
    \end{tabularx}
    \begin{tablenotes}[flushleft]
      \footnotesize
      \item[] Bold labels indicate the eight variables shown in the main text. QSOIL includes ground evaporation and sublimation minus dew. Water and carbon fluxes are converted from per-second to per-day rates. For TSOI, H2OSOI, SOILLIQ and SOILICE, the six-hourly metrics use all valid soil layers, whereas the mean fields and scatter plots use the first soil layer.
    \end{tablenotes}
  \end{threeparttable}
\end{table}